\documentclass[11pt]{article}

\usepackage{latexsym}
\usepackage{amssymb}
\usepackage[usenames,dvipsnames]{color}

\def\een{\\ &=&}
\def\ua{$U(1)_a$}
\usepackage{framed}
\usepackage{multirow}
\usepackage{bbm}
\def\thefootnote{\fnsymbol{footnote}}

\DeclareMathAlphabet   {\mathsc}{OT1}{cmr}{m}{sc}

\newcommand{\myref}[1]{(\ref{#1})}
\newcommand{\mysec}[1]{Section~\ref{#1}}
\newcommand{\myapp}[1]{Appendix~\ref{#1}}

\def\Del{\Delta}
\def\half{{1\over2}}
\def\bea{\begin{eqnarray}}
\def\eea{\end{eqnarray}}
\def\beq{\begin{equation}}
\def\eeq{\end{equation}}

\def\bL{\bar{\Lambda}}
\def\ux{$U(1)_X$}

\def\uo{$U(1)$}
\def\dx{\delta_X}
\def\vx{V_X}

\def\superint{\int d^{4}\theta}

\newcommand{\WaWa}{ W_a^{\alpha} W^a_{\alpha}}

\newcommand{\Da}{{\cal D}_{\alpha}}
\newcommand{\Db}{{\cal D}^{\dot{\beta}}}

\newcommand{\Wa}{W_{\alpha}}

\newcommand{\Xa}{X_{\alpha}}

\newcommand{\Ga}{G_{\alpha}}
\newcommand{\Xc}{X^{\alpha}}

\newcommand{\fa}{f_{\alpha}}
\newcommand{\fc}{f^{\alpha}}
\newcommand{\ga}{g_{\alpha}}
\newcommand{\gc}{g^{\alpha}}

\def\im{{\rm Im}}

\def\tg{\tilde{g}}

\def\G{{\cal G}}

\def\tG{{\widetilde G}}

\def\tF{{\tilde F}}

\def\dZ{{\dot Z}}
\def\dY{{\dot Y}}

\def\D{{\cal D}}

\def\bD{\bar{\D}}

\def\pp{\partial}

\def\ibar{\bar{\imath}}

\def\[{\left [}
\def\]{\right ]}
\def\({\left (}
\def\){\right )}
\def\lbr{\left\{}
\def\rbr{\right\}}
\def\r{\right|}
\def\l{\left.}

\def\T{\bar{T}}

\def\z{\bar{z}}
\def\q{\bar{q}}
\def\del{\delta}

\def\deta{{\dot\eta}}

\def\S{{\bar{S}}}

\def\Tr{{\rm Tr}}

\def\Ti^{T^{(i)}}

\def\f{\bar{f}}
\def\bR{\bar{R}}

\def\L{{\cal L}}

\def\t{\bar{t}}

\def\m{\bar{m}}

\def\cM{{\cal{M}}}

\def\Z{{\bar{Z}}}

\def\bPh{\bar{\Phi}}

\def\bF{\bar{F}}
\def\Q{\bar{Q}}

\def\fb{\bar{f}}

\def\a0{\alpha_0}
\def\chiproj{(\bD^2 - 8R)}

\def\eee{\nonumber \\ &=&}
\def\ddd{\nonumber \\ &&}
\def\mmm{\nonumber \\ &&\qquad}
\def\nnn{\nonumber \\ }
\def\hc{ + {\rm h.c.}}

\begin{document}
\newcommand\text[1]{{\rm {#1}}}

\def\subsubsubsection{\paragraph}

\begin{titlepage}
\begin{center}

\hfill August 2019\\[1in]

{\large {\bf ANOMALY CANCELLATION IN EFFECTIVE SUPERGRAVITY 
    FROM THE HETEROTIC STRING WITH AN ANOMALOUS \uo}}\footnote{This
  work was supported in part by the Director, Office of Science,
  Office of High Energy and Nuclear Physics, Division of High Energy
  Physics, of the U.S. Department of Energy under Contract
  DE-AC02-05CH11231 and in part by the National
  Science Foundation under grants PHY-1316783.}  \\[.1in]

 Mary K. Gaillard and Jacob M. Leedom\\[.1in]

{\em Department of Physics and Theoretical Physics Group,
 Lawrence Berkeley Laboratory, \\
 University of California, Berkeley, California 94720}\\[1in] 

\end{center}

\begin{abstract}

We show that a choice of Pauli-Villars regulators allows the
cancellation of all the conformal and chiral anomalies in an effective
field theory from $\mathbb{Z}_3$ compactification of the heterotic
string with two Wilson lines and an anomalous \uo.
  
\end{abstract}
\end{titlepage}

\newpage

\renewcommand{\thepage}{\roman{page}}
\setcounter{page}{2}
\mbox{ }

\begin{center}
{\bf Disclaimer}
\end{center}

\vskip .2in

\begin{scriptsize}
\begin{quotation}
  This document was prepared as an account of work sponsored by the
  United States Government. While this document is believed to contain
  correct information, neither the United States Government nor any
  agency thereof, nor The Regents of the University of California, nor
  any of their employees, makes any warranty, express or implied, or
  assumes any legal liability or responsibility for the accuracy,
  completeness, or usefulness of any information, apparatus, product,
  or process disclosed, or represents that its use would not infringe
  privately owned rights.  Reference herein to any specific commercial
  products process, or service by its trade name, trademark,
  manufacturer, or otherwise, does not necessarily constitute or imply
  its endorsement, recommendation, or favoring by the United States
  Government or any agency thereof, or The Regents of the University
  of California.  The views and opinions of authors expressed herein
  do not necessarily state or reflect those of the United States
  Government or any agency thereof, or The Regents of the University
  of California.
\end{quotation}
\end{scriptsize}

\vskip 2in

\begin{center}
\begin{small}
{\it Lawrence Berkeley Laboratory is an equal opportunity employer.}
\end{small}
\end{center}

\newpage
\renewcommand{\theequation}{\arabic{section}.\arabic{equation}}
\renewcommand{\thepage}{\arabic{page}}
\setcounter{page}{1}
\def\thefootnote{\arabic{footnote}}
\setcounter{footnote}{0}

\section{Introduction}\label{intro}
Starting with the determination of the full anomaly structure of
Pauli-Villars (PV) regularized supergravity~\cite{bg}, we recently
showed~\cite{gl} that an appropriate choice of PV regulator fields
allows for cancellation of all the T-duality (hereafter referred to as
``modular'') anomalies by the four-dimensional version of the
Green-Schwarz term in $\mathbb{Z}_3$ and $\mathbb{Z}_7$
compactifications of the heterotic string without Wilson
lines.\footnote{Corrections to this paper are given in \myapp{cors}.}
We further matched our results to a string calculation~\cite{ss} of
the chiral anomaly in those theories.  Here we extend our results to a
specific $\mathbb{Z}_3$ compactification~\cite{fiqs} (hereafter referred
to as FIQS) with two Wilson
lines and therefore an anomalous \uo, hereafter referred to as \ux.
In the following section we briefly describe the orbifold model we are
studying. In Section 3 we outline the four-dimensional Green-Schwarz
mechanism and the structure of the anomaly when an anomalous \uo\, is
present.  In Section 4 we discuss some aspects of the cancellation of
ultra-violet (UV) divergences and anomaly matching that are specific
to the case with an anomalous \uo, as well as some simplifications
with respect to the $\mathbb{Z}_7$ case studied in~\cite{gl}.  We
summarize our results in Section 5.  The full set of conditions for
cancellation of UV divergences and anomaly matching are given in
\myapp{conds}{, a sample solution to these constraints is presented in
\myapp{sols}, and the full spectrum for the FIQS model is displayed
in \myapp{charges}}.  The determination of the correct Pauli-Villars (PV)
masses can have implications for soft supersymmetry breaking
terms~\cite{soft}.

\section{The FIQS model}\label{fiqs}
\setcounter{equation}{0}

Here we will give a brief review of the orbifold model we will
consider for the rest of the paper. The FIQS model~\cite{fiqs} is a $\mathbb{Z}_3$
orbifold compactification of the 10d $E_8 \otimes E_8$ heterotic
string compactified to $T^6$ with two Wilson lines and a nonstandard
embedding for the shift vector. The embeddings of the shift vector and
Wilson lines are given by
\noindent
\bea
		V &=&  \frac{1}{3} (1,1,1,1,2,0,0,0)(2,0,0,0,0,0,0,0)^\prime\\
		a_1 &=&  \frac{1}{3} (0,0,0,0,0,0,0,2)(0,1,1,0,0,0,0,0)^\prime\\
		a_3 &=&  \frac{1}{3} (1,1,1,2,1,0,0,1,1)(1,1,0,0,0,0,0,0)^\prime
\eea
\noindent
Where the prime indicates that the last 8 elements of the above
vectors correspond to the second factor of $E_8$.  With these
specifications, the massless spectrum of the FIQS model can be worked
out following the standard recipes~\cite{refs1}. The 4D gauge group is
$SU(3)\otimes SU(2) \otimes SO(10) \otimes U(1)^8$. The generators of
the eight $U(1)$ factors can be written as linear combinations of the
$E_8 \otimes E_8$ Cartan subalgebra generators $H^I$ as
\noindent
\bea
Q_a = \sum_{I=1}^{16}q_a^I H^I \eea
\noindent
The constants $q_a^I$ are determined by requiring that $q_a\cdot q_b =
0$ and $q_a\cdot \alpha_{bj} = 0$, where the $\alpha_{bj}$ are the
sixteen dimensional simple root vectors of the nonabelian gauge group
factors. Thus the index b corresponds to $SU(3)$, $SU(2)$, or $SO(10)$
and $j$ runs over the rank of each group. One choice of $q_a$'s
is~\cite{refs2}:

\bea
		\vec q_1 &=&  6(1,1,1,0,0,0,0,0)(0,0,0,0,0,0,0,0)^\prime\\
		\vec q_2 &=&  6(0,0,0,1,-1,0,0,0)(0,0,0,0,0,0,0,0)^\prime\\
		\vec q_3 &=&  6(0,0,0,0,0,1,0,0)(0,0,0,0,0,0,0,0)^\prime\\
		\vec q_4 &=&  6(0,0,0,0,0,0,1,0)(0,0,0,0,0,0,0,0)^\prime\\
		\vec q_5 &=&  6(0,0,0,0,0,0,0,1)(0,0,0,0,0,0,0,0)^\prime\\
		\vec q_6 &=&  6(0,0,0,0,0,0,0,0)(1,0,0,0,0,0,0,0)^\prime\\
		\vec q_7 &=&  6(0,0,0,0,0,0,0,0)(0,1,0,0,0,0,0,0)^\prime\\
		\vec q_8 &=&  6(0,0,0,0,0,0,0,0)(0,0,1,0,0,0,0,0)^\prime
\eea
To get the charges of the matter fields, one normalizes the $U(1)_a$ generators as
\bea
	 Q_a \rightarrow \frac{1}{\sqrt{2}\left|q_a\r}Q_a,
\eea
\noindent
where the $\sqrt{2}$ is inserted to adhere to the standard
phenomenological normalization. For this choice, one finds that the
traces $Q_6$, $Q_7$, and $Q_8$ are all nonzero. One can perform a
re-definition of the generators so that only one factor of U(1) has a
nonzero trace. In~\cite{fiqs}, the following re-definition was made:
\noindent
\bea
		q^{(FIQS)}_6 &=&  q_6+q_7\\
		q^{(FIQS)}_7 &=&  q_7+q_8\\
		q_X &=&  q_6 - q_7 +q_8
\eea
\noindent
While $\Tr\[Q_6^{(FIQS)}\] = \Tr\[Q_7^{(FIQS)}\] = 0 $ in this basis,
one also has $\Tr\[Q^{(FIQS)}_6 Q^{(FIQS)}_7 Q_X\] \neq 0$ which is
rather undesirable. Therefore, we will use a different choice such
that the above mixed anomaly does not appear. In particular, we define
\noindent
\bea
q^{(N)}_6 &=& q_6 - q_8 =q^{(FIQS)}_6 - q^{(FIQS)}_7
\\ q^{(N)}_7 &=& q_6 + 2 q_7 + q_8 = q^{(FIQS)}_6 + q^{(FIQS)}_7
\eea
\noindent
In what follows, we will simply drop the superscript N and use these as the definition of the $U(1)_6$ and $U(1)_7$ generators. As a final note, the charges defined above are generally not orthogonal to one another, i.e. $\Tr\[Q_a Q_b\] \neq 0$
for some $a\neq b$. It is possible to define a new set of charges that are mostly orthogonal to one another, but we will not need to do so for our purposes.

We close this section with some relations among the gauge charges
$q^p_a$ and modular weights $q^p_n$ of the chiral superfields
$\Phi^p$ of the model.  These will be useful in the analysis that
follows.  These include the universality conditions
\bea 8\pi^2b &=& C_a + \sum_p\(2q^p_i -  1\)C^p_a = 
{1\over24}\(2\sum_p q^p_n - N + N_G - 21\)\quad\forall\quad i,a,\nnn
- 2\pi^2\dx &=&  {1\over24}\Tr T_X =  {1\over3}\Tr T_X^3 = 
 \Tr(T^2_aT_X)\quad \forall\quad a\ne X\label{uxconds}.\eea
Here $C_a$ is the quadratic Casimir in the adjoint representation of
the gauge group factor $\G_a$ and $C_a^p$ is the Casimir for the
representation of the chiral supermultiplet $\Phi^p$, $T_a$ is a 
generator of $\G_a$, and $N,N_G$ are the number of chiral and gauge
supermultiplets respectively, with, in the FIQS model,
\beq N = 415, \qquad N_G = 64,\qquad 8\pi^2b = 6,\qquad - 4\pi^2\dx =
3\sqrt{6}.\label{Nbdx}\eeq
\noindent
In addition we will use the sum rules
\noindent 
\bea \sum_p q^p_n &=& A_1, \qquad \sum_p q^p_m q^p_n = A_2 +
B_2\del_{m n},\nnn \sum_p q^l q^p_m q^p_n &=& A_3 + B_3\(\del_{l m} +
\del_{m n} + \del_{n l}\) + C_3\del_{l m}\del_{m n}, \nnn
\sum_b q^b_a q^b_n &=& Q_{1a},\qquad \sum_b q^b_a q^b_m q^b_n =
Q_{2a} + P_{2a}\del_{m n},\label{qsums}\eea
\noindent
with, in particular,
\beq B_2 = 42,\qquad  P_{2X} = 5\sqrt{6}.\label{b2psx}\eeq

\section{Anomalies and anomaly cancellation with an anomalous \uo}
\label{anom}
\setcounter{equation}{0}
The effective supergravity theory from generic orbifold compactifications
with Wilson lines is anomalous under both \ux\, and T-duality:
\bea T'^i &=& {a_i - ib_i T^i\over i c_i T^i + d_i}, \qquad a_i b_i 
- c_i d_i = 1,
\qquad a_i,b_i,c_i,d_i \in {\bf Z},\qquad i = 1,2,3,\nnn
\Phi'^a &=& e^{-\sum_i q^a_i F^i(T^i)}\Phi^a,\qquad F^i(T^i) = 
\ln(i c_i T^i + d_i),\label{modtr}\eea
where $\Phi^a$ is any chiral supermultiplet other than a diagonal
K\"ahler modulus $T^i$, and $q^a_i$ are its modular weights.  

We are working in the covariant superspace formalism of ref.~\cite{bgg}
in which the chiral multiplets $Z^p = T^i,S,\Phi^a$, with $S$ the
dilaton superfield, are {\it covariantly} chiral:
\beq \Db Z^p = 0,\label{deriv}\eeq
\noindent
with $\D_A,\;A = a,\alpha$ a fully covariant superspace derivative.
In particular, under a \uo\, gauge transformation
\noindent
\beq Z'^p = g^{q^p_a}Z^p,\qquad \Z'^p = g^{-q^p_a}\Z^p,\qquad
A'^a_A = A_A^a - g^{-1}\D_A g,\label{gauge}\eeq
\noindent
where $g$ is a hermetian superfield, and $A_A$ is the gauge potential
in superspace. Gauge invariance assures that holomorphy of the
superfield is maintained under \myref{gauge}.  If gauge invariance is
unbroken, the gauge potential $A_A$ does not appear explicitly in the
superspace Lagrangian. Instead the usual Yang-Mills superfield
strength $\Wa$ is obtained as a component of the two-form superfield
strength $F_{A B}$.  One can still introduce~\cite{bgg} a superfield
superpotential $V_a$ such that
\noindent
\beq \Wa = - {1\over8}\chiproj\Da V_a, \qquad V'_a = V_a + \Lambda_a
+\bar\Lambda_a,\label{wava}\eeq
\noindent 
but $V_a$ never appears in the Lagrangian and the chiral superfield
$\Lambda_a$ is independent of $g$ in \myref{gauge}.

However in the presence of an anomalous \uo, gauge invariance is
broken. It is easy to see that the UV divergences cannot be regulated
by PV fields that all have \ux\, invariant masses.  There is a 
quadratically divergent term proportional to $D_X\Tr T_X$, where $D_X$ 
is the auxiliary field of the \ux\, supermultiplet, which
must be cancelled by the analogous term from the PV sector. Invariant
masses require the coupling of PV fields with equal and opposite
charges that do not contribute to $(\Tr T_X)_{PV}.$  Noninvariant
masses arise from the superpotential for PV fields $\Phi^C$:
\noindent
\beq W(\Phi^C,\Phi'^C) = \mu_C\Phi^C\Phi'^C, \label{wphic}\eeq
\noindent
with $\mu_C$ constant (in the absence of threshold corrections, as for
the cases considered here).  If $Q_X^C +Q'^C_X\ne 0$, holomorphy of
\myref{wphic} is not respected under \myref{gauge} for $a =X$.  For
this reason we do not include the \ux\, connection in the covariant
derivative \myref{deriv}.  Instead of \myref{gauge} we require
\noindent
\beq \Phi'^C = e^{- Q^C_X\Lambda}\Phi^C,\qquad \bPh'^C =
e^{-Q^C_X\bL}\bPh^C\label{phiux}\eeq
\noindent
under a \ux\, transformation, and the K\"ahler potential depends on
\ux-charged fields through the invariant operators $\bPh e^{Q_X\vx}\Phi$.

It was shown in~\cite{bg} that modular noninvariant masses can be
restricted to a subset of PV chiral supermultiplets $\Phi^C$ with
diagonal K\"ahler metric: 
\beq K(\Phi^C,\bar\Phi^C) = \exp[f^C(Z,\Z)]|\Phi^C|^2.\label{kphic}\eeq
\noindent
and superpotential \myref{wphic}.

As in~\cite{gl}, we define a superfield
\beq \cM^2_C = \cM^2_{C'} = \exp(K - f^C - f'^C) = \exp(K - 2\f^C),
\qquad \f^C = \half(f^C + f'^C),\label{M2}\eeq
\noindent
whose lowest component $m_C^2 = \l\cM^2_C\r$ is the $\Phi^C,\Phi'^C$
squared mass. Then the anomalous part of the one-loop corrected
supergravity Lagrangian takes the form~\cite{bg}
\bea \L_{\rm anom} &=& \L_0 + \L_1 + \L_r = \superint E\(L_0 + L_1+
L_r\) \equiv \superint E\Omega,\label{SFtotanom}\eea
where $E$ is the superdeterminant of the supervielbein, and
\bea L_0 &=& {1\over8\pi^2}\[\Tr\eta\ln\cM^2\Omega_0 + K\(\Omega_{G B} 
+ \Omega_D\)\],\label{totomega}\eea
\noindent
with $\eta = \pm 1$ the PV signature.  The operators in
\myref{totomega} are given explicitly in~\cite{bg,gl}, except that now
\beq \Omega_0 = \Omega^0_{\rm Y M} + \Omega'_0, \label{Omega0}\eeq
\noindent
where $\Omega'_0$ contains the Gauss-Bonnet Chern-Simons superfield and
operators composed of auxiliary superfields of the gravity
supermultiplet, and
\noindent 
\beq \Omega^0_{\rm Y M} = \sum_{a\ne X}\Omega^a_{\rm YM} =
\Omega_{\rm YM} - \Omega^X_{\rm YM},
\label{omegaym}\eeq
\noindent
is the Yang-Mills Chern-Simons superfield without the \ux\, term, and
and $\Omega^a_{\rm Y M}$ is defined by its chiral projection:
\beq \chiproj\Omega^a_{\rm Y M} = \WaWa.\eeq
\noindent
$\Omega_r$ is composed of terms linear and higher order in $\ln\cM$,
and $\Omega_D$ represents a ``D-term'' anomaly~\cite{bg,gl} that,
together with a contribution to the Gauss-Bonnet term $\Omega_{\rm G
  B}$, arises from uncanceled total derivatives with logarithmically
divergent coefficients, requiring the introduction of a
field-dependent cut-off:
\beq \pp_\mu\Lambda = {1\over4}\pp_\mu K.\label{lambda} \eeq
\noindent
$L_1$ is defined by its variation:
\beq \Del L_1 = {1\over8\pi^2}{1\over192}\Tr\eta\Del\ln\cM^2\Omega'_L 
= {1\over8\pi^2}{1\over192}\Tr\eta H\Omega'_L\hc,\label{dell1}\eeq
\noindent
where under \myref{modtr} and \myref{phiux} $\ln\cM^2$ transforms as
\beq \Del\ln\cM^2 = H + \bar H,\label{DelM2}\eeq
with $H$ holomorphic. Defining 
\bea \chiproj\Omega_f &=& \fc\fa,\qquad \chiproj\Omega_{\f} = \bar\fc
\bar\fa,\qquad \chiproj\Omega_{\f X} = \bar\fc\Xa,\nnn
\fa &=& - {1\over8}\chiproj\Da f,\qquad
\bar\fa = - {1\over8}\chiproj\Da\f,\label{Omegaf}\eea
we have
\bea \Omega'_L &=& 192\Omega_f - 128\Omega_{\f} - 64\Omega_{\f X},\nnn
\Del L_1 &=& {1\over8\pi^2}\Tr\eta H\(\Omega_f - {2\over3}\Omega_{\f} -
{1\over3}\Omega_{\f X}\)\hc\label{Omegalprime}\eea
\noindent
In the presence of an anomalous \ux\, the form of $f^C$ is taken to be
\bea f^C &=& \alpha^C K(Z,\Z) + \beta^C g(T,\T) + \delta^C k(S,\S)
+ \sum_n q^C_n g^n(T^n,\T^n) + Q^C_X V_X,\nnn 
\f^C &=& \bar\alpha^C K + \bar\beta^C g
+ \bar\delta^C k + \sum_n\q^C_n g^n + \Q^C_X V_X,\nnn 
H^C &=& \(1 - 2\bar\gamma^C\)F(T) - 2\sum\q^C_n F^n(T^n)
- 2\Q^C_X\Lambda,
\qquad \bar\gamma^C = \bar\alpha^C + \bar\beta^C,\label{deffh}\eea
\noindent
{ where $k$ is the dilaton k\"ahler potential, and $g$ is defined
in \myref{kmod} below.}
The traces in $\Del\L_{\rm anom}$ can be evaluated using only PV
fields with noninvariant masses or using the full set of PV fields,
since those with invariant masses, $H^C=0$, drop out.  The
contribution $\Del L_0$ to the anomaly is linear in the parameters
$\alpha^C,\beta^C, q^C_n,Q^C_X$, and the trace of the coefficient of
$\Omega'_0$ is completely determined by the sum rules~\cite{pv}
\bea N' &=& \sum_C\eta^C = - N - 29,\qquad N'_G = 
\sum_\gamma\eta_\gamma^V = - 12 - N_G,\nnn
\sum_C\eta^C f^C &=& - 10K - \sum_p q^p_n g^n - \sum_a q^a_X V_X,
\label{sums}\eea
\noindent
that are required to assure the cancellation of quadratic and
logarithmic divergences.  In \myref{sums} the index $C$ denotes any
chiral PV field, the index $\gamma$ runs over the Abelian gauge PV
superfields that are needed to cancel some gravitational and
dilaton-gauge couplings, and the sum over $p$ includes all the light
chiral multiplet modular weights with $q^S_n = 0,\; q^{T^i}_n =
2\del^i_n$.  All PV fields with noninvariant masses have $\del = 0$,
and most\footnote{There is a set of chiral multiplets in the adjoint
  representation of the gauge group that has $f = K - k$; these get
  modular invariant masses though their coupling in the superpotential
  to a second set with $f = k$. These cancel renormalizable gauge
  interactions and gauge-gravity interactions, respectively.  Together
  with a third set, that has $f = 0$ and contributes to the anomaly,
  they cancel the Yang-Mills contribution to the
  beta-function.\label{foot3}} with $\del\ne0$ have $\alpha = \beta =
q_n = 0 = Q^C_X$.  For the purposes of the present analysis we can
largely ignore the latter.  Similarly, the cancellation of linear
divergences that give rise to the chiral anomaly proportional to
\noindent
\beq \im\Tr\phi G\cdot\tG\ni\im\half\sum_{a\ne X}\lbr F(t)C_a - 
\sum_p\[F(t) - 2\sum_n q^p_n F^n(t^n) - 2 q^p_X\lambda\]
(T^p_a)^2\rbr F^a\cdot\tF_a\label{gaugeanom}\eeq
\noindent
fixes the coefficient of $\Omega^0_{Y M}$.  Here $G_{\mu\nu}\ni -i
T_a F^a_{\mu\nu}$ is the field strength associated with the fermion
connection, $t^i = \l T^i\r$, $\lambda = \l\Lambda\r$ are the lowest
components of the chiral supermultiplets $T^i,\Lambda$, and a
left-handed fermion $f$ transforms as
\noindent
\beq f \to e^\phi f \label {defphi}\eeq
\noindent under modular and \ux\, transformations; $\phi = -{i\over2}
\im F$ for gauginos, and 
\noindent 
\beq \phi = {i\over2}\im F - \sum_n q^p_n F^n(t^n) - q^p_X\lambda
\label{phichi}\eeq
for chiral fermions $\chi^p$.  The compensating PV contribution
\noindent
\beq \im\(\Tr\eta\phi G\cdot\tG\)_{P V}\ni\im\sum_C\eta^C\(\phi^C + 
\phi'^C\)(T^C_a)^2F_a\tF^a = - \im\Tr\phi G\cdot\tG \label{PVgauge}\eeq
\noindent 
that cancels \myref{gaugeanom} determines the anomaly coefficient of
$\Omega^0_{Y M}$, since for each pair $\Phi^C,\Phi'^C$ the sum of
fermion phases $\phi^C + \phi'^C = H^C$ is just the holomorphic part
of the variation \myref{DelM2}, \myref{deffh} of the PV mass term
$\Del\ln\cM^2_C$.

In the chiral formulation for the dilaton, the anomaly is cancelled by
the variation of the superspace Lagrangian
\beq \L =  \superint E\(S + \S\)\Omega.\,
\label{superl}\eeq
\noindent
where $\Omega$ is the real superfield introduced in \myref{SFtotanom}.
The quantum Lagrangian varies according to
\beq\Del\L_{\rm anom} = \superint\lbr b\[F(T) + \bF(\T)\]
- {\del_X\over2}\(\Lambda + \bL\)\rbr\Omega,\label{delL}\eeq
so the full Lagrangian is invariant provided 
\beq\Del S = - b F(T) + {\del_X\over2}\Lambda, \qquad
F = \sum_i F^i.\label{delS}\eeq
However the classical K\"ahler potential for the dilaton is no longer
invariant and must be modified:
\beq k_{\rm class}(S,\S) = -\ln(S + \S) \to k(S,\S) =  -\ln(S + \S
+ V_{G S}), \label{kdil}\eeq
\noindent 
where $V_{G S}$ is a real function of $\vx$ and of the chiral
supermultiplets; it transforms under \myref{modtr} and \myref{wava},
\myref{phiux} as
\beq \Del V_{G S} = b\(F + \bF\) - {\del_X\over2}\(\Lambda + \bL\).
\label{delV}\eeq
A simple solution consistent with string calculation 
results~\cite{prev,dterm} 
is
\beq V_{G S}  = b g(T,\T) - {\del_X\over2}V_X,\label{defV}\eeq
where
\beq g(T,\T) = \sum_i g^i(T^i,\T^i),\qquad g^i = - \ln(T^i + \T^i)
\label{kmod} \eeq
\noindent
is the K\"ahler potential for the moduli.  The modification
\myref{kdil} is the 4d Green-Schwarz (GS) term in the chiral
formulation.  As discussed in~\cite{gl}, the 4d GS mechanism is more
simply formulated in the linear multiplet formalism~\cite{bgg} for the
dilaton. In this case the linear dilaton superfield $L$ remains
invariant, its K\"ahler potential is unchanged, and instead one adds a
term to the Lagrangian:
\bea \L_{G S} &=& - \superint E L V_{G S},\qquad
\Del\L_{G S} =  - \Del\L_{\rm anom}\label{lings}
\eea
Only terms in the anomaly that are linear in the
combination $\tilde H$, where
\beq \tilde H = b F(T) - {\del_X\over2}\Lambda,\label{tildeH}\eeq
\noindent
can be canceled by the Green-Schwarz term.  The values of $b$ and
$\dx$ are fixed by the conditions \myref{sums}, \myref{PVgauge} for
the cancellation of divergences, together with the universality
conditions \myref{uxconds}, that hold for all $\mathbb{Z}_3$ and $\mathbb{Z}_7$ orbifold
compactifications.

In contrast to $\L_0$, the contributions to the anomaly from $\L_1$
and $\L_r$ are nonlinear in the parameters $\alpha,\beta,q_n,Q_X$, and
depend on the details of the PV sector. In particular $\L_r$ has no
terms linear in $\ln\cM$ and must vanish. To insure that the anomaly
coefficient depends on the T-moduli only through $F(T)$ we
impose~\cite{gl}
\noindent
 \beq\q^C_n = 0\label{qbarcond}\eeq 
\noindent 
for (almost\footnote{The exception is for some PV fields,
    introduced in \myapp{abel}, needed to cancel divergences from light
    fields with Abelian gauge charges.}) all PV fields with
noninvariant masses.

\section{The anomaly and cancellation of UV divergences in the FIQS 
model}\label{fiqsanom}
\setcounter{equation}{0}

The full set of conditions for cancellation of the divergences and for
obtaining an anomaly linear in $\tilde H$, Eq. \myref{tildeH}, that
matches the string result~\cite{ss} is given in the \myapp{conds}.  In this section
we outline some features of the case of $\mathbb{Z}_3$ with an anomalous \ux.
We will be primarily concerned with the contribution
of $\Del L_1$, Eq. \myref{Omegalprime}, to the anomaly.  This
expression is nonlinear in the parameters $q^C_n,Q^C_X$ of the PV
fields, and therefore model dependent, as noted above.  This was
illustrated in~\cite{gl} where it was shown that cancellation of the
modular anomaly requires \myref{qbarcond}. However, the contribution
cubic in $Q^C_X$ is model independent. It is given by
\noindent 
\beq \Del L_1(Q_X^3) = - {2(\Lambda + \bL)\over24\pi^2}\Tr\eta\bar Q_X\(3Q_X^2 -
2\bar Q^2_X\)\Omega_{YM}^X = 
- {2(\Lambda + \bL)\over24\pi^2}\Tr\eta Q_X^3\Omega_{YM}^X,\label{l1Q3}\eeq
\noindent
where the sum is over all PV fields, and we used the definition (3.6),
\myref{deffh} of $\bar Q^X$ and the fact that
\beq \sum_C\eta^C(Q_X^C)^p(Q_X'^C)^{p'} = 
\sum_C\eta^C(Q_X^C)^{p'}(Q_X'^C)^p,\eeq
\noindent
for any powers $p,p'$.  Cancellation of the term in $\Tr\phi
G\cdot\tG$ that is cubic in $Q_X^3$ requires
\beq -{2(\Lambda + \bL)\over24\pi^2}\Tr\(\eta Q^3_X\)\Omega_{YM}^X = 
{2(\Lambda + \bL)\over24\pi^2}\Tr\(q^3_X\)\Omega_{YM}^X 
= - {\dx\over2}(\Lambda + \bL)\Omega_{YM}^X,\label{q3}\eeq
\noindent 
from \myref{uxconds}, so the anomaly \myref{l1Q3} is consistent with
the requirement for anomaly cancellation.

In contrast, anomaly terms quadratic in $Q_X^2$ are model dependent.
For example, in~\cite{bg} it was assumed that $\fb^C = f^C$ for 
all PV fields with noninvariant masses, giving a contribution
\noindent
\bea \Del L_1(F Q_X^2) &=& {F + \bF\over24\pi^2}\Tr\eta\(1 - 2\bar\gamma\)
\(3Q_X^2 - 2\bar Q^2_X\)\Omega_{YM}^X \label{l1Q2F}\een   
{F + \bF\over24\pi^2}\Tr\eta\(1 - 2\bar\gamma\) Q_X^2\Omega_{YM}^X =
{F + \bF\over24\pi^2}\Tr q_X^2\Omega_{YM}^X = {b\over3}(F + \bF)\Omega_{YM}^X,
\label{l1Q2F2}\eea
\noindent
from \myref{gaugeanom} and \myref{PVgauge} with $a=X$, and
\myref{uxconds}.  Here we instead assume, in addition to
\myref{qbarcond}, that $\bar Q_X^C = 0$ if $1 - \bar\gamma\ne0$, that
is PV masses can be noninvariant under either T-duality or \ux, but
not both.  In this case the last term in \myref{l1Q2F} drops out and we
recover a factor three, in agreement with the requirement for anomaly
cancellation.

The full set of PV fields sufficient to regulate light field couplings
is described in Section 3 of~\cite{bg}.  These include a set $\dZ^P =
\dZ^I,\dZ^A$, with negative signature, $\eta^{\dZ} = -1,$ that
regulates most of the couplings, including all renormalizable
couplings, of the light chiral supermultiplets $Z^p = T^i,\Phi^a$.
The $\dZ$ get invariant masses through a superpotential coupling to PV
fields $\dY_P$ with the same signature, opposite gauge charges and the
inverse K\"ahler metric:
\beq (T_a)_{\dY} = - (T_a^T)_{\dZ} = - (T_a^T)_Z.\label{Ta}\eeq
\noindent
It remains to cancel the divergences introduced by the fields $\dY$.
To this end we take the following set:
\bea \psi^{P n}: && f^{P n} = \alpha_\psi^P K + \beta_\psi^P g +
q_\psi^P g^n + Q_\psi^P V_X, \qquad \alpha_\psi^P + \beta_\psi^P 
= \gamma_\psi^P,\qquad \bar q_\psi^P = 0,\nnn 
T^P: && f_T^P = \alpha_T^P K + \beta_T^P g + Q_T^P\vx, 
\qquad \alpha_T^P + \beta_T^P = \gamma_T^P,\nnn
\phi^C: && f^{\phi^C} = {\alpha^C K}.\label{pvfields}\eea 
\noindent 
{In the solution to the constraints given in \myapp{sols}, the
  $\psi^C$ and $T^C$ are further subdivided, together with additional
fields, into sets $S_a,\; a
  = 1,\ldots,12$, some of which are charged under the nonanomalous
gauge group.}  The $\phi^C$ regulate certain gravity
supermultiplet loops and nonrenormalizable coupling of chiral
multiplets.  These must be included together with the other PV fields
introduced above in implementing the sum rules \myref{sums}.  Their
contributions will be included in all the finiteness and anomaly
conditions that involve only the parameters $\alpha$ in
\myref{pvfields}; otherwise they play no role in the analysis below.
In the expressions given in the remainder of this section, we drop
terms that contain only $\Xa$ or $X_{\mu\nu}$ since their
contributions are included in the sums \myref{sums} and the additional
sum rule~\cite{pv}
\beq \sum_C\eta^C\alpha_C^2 = - 4\label{a2rule}.\eeq

In~\cite{gl} we also introduced pairs $\Phi^P,\Phi'^P$ with modular 
invariant masses that did not contribute to the anomaly, but played 
an important role in canceling certain divergences. 
However, because the $\mathbb{Z}_3$ sum rules \myref{qsums} are much simpler
than the analogous sum rules for the $\mathbb{Z}_7$ case studied in~\cite{gl},
here we need only the set in \myref{pvfields}.

The quadratic and logarithmic divergences we are concerned with here
involve the superfield strengths $-i(T_a)\Wa^a$,
\beq \Gamma^C_{D\alpha} = - {1\over8}\chiproj\Da Z^p\Gamma^C_{D p},
\label{gamma}\eeq
\noindent
and
\beq X_\alpha = - {1\over8}\chiproj\Da K, \eeq
\noindent
associated with the Yang-Mills, reparameterization and K\"ahler 
connections, $i(T_a)^C_D A_\mu,\; \pp_\mu Z^p\Gamma^C_{p D}$
and $\del^C_D\Gamma_\mu$, respectively, where
\beq \Gamma_\mu = {i\over4}\(\D_\mu z^i K_i - \D_\mu\z^{\m}K_{\m}\).  
\label{Gamma}\eeq
\noindent

Cancellation of quadratic divergences requires
\beq \Tr\eta\Gamma_\alpha = \Tr\eta T_X = 0,\label{quad}\eeq
\noindent
and cancellation of logarithmic divergences requires 
\beq \Tr\eta\Gamma_\alpha\Gamma_\beta = \Tr\eta\Gamma_\alpha T^a =
\Tr\eta(T^a)^2 = 0,\label{log}\eeq
\noindent
where $\eta = + 1$ for light fields.  Cancellation of all
contributions linear and quadratic in $\Xa$ is assured by the
conditions in \myref{sums} and \myref{a2rule}. 
\noindent
The Yang-Mills contribution to the term quadratic in $\Wa$ is canceled
by chiral fields in the adjoint (see footnote on page~\pageref{foot3})
that we need not consider here.  Finally, cancellation of linear
divergences requires cancellation of the imaginary part of
\beq \Tr\eta X_\chi = \Tr\eta\phi G\cdot\tG, \qquad \tG^{\mu\nu}= \half
\epsilon^{\mu\nu\rho\sigma} G_{\rho\sigma},\label{defX}\eeq
\noindent
where $G_{\mu\nu}$ is the field strength associated with the fermion
connection;\footnote{Here we neglect the spin connection whose
contribution was discussed in~\cite{gl}.} for left-handed fermions:
\beq G_{\mu\nu} = - \Gamma^C_{D\mu\nu} + iF^a_{\mu\nu}(T_a)^C_D
+ \half X_{\mu\nu}\del^C_D,\label{defG}\eeq
\noindent
where 
\bea X_{\mu\nu} &=& \(\D_\mu z^i\D_\nu\z^{\m} - 
\D_\nu z^i\D_\mu\z^{\m}\)K_{i\m} - iF^a_{\mu\nu}(T_a z^i)K_i \eee
2i\(\pp_\mu\Gamma_\nu - \pp_\nu\Gamma_\mu\),\qquad i = p,s,
\label{xmunu}\eea
\noindent 
is the field strength associated with the K\"ahler connection
\myref{Gamma}.  For a generic PV superfield $\Phi^C$ with diagonal
metric, its fermion component $\chi^C$ transforms under \myref{modtr}
and \myref{phiux} as
\beq \chi'^C = e^{\phi^C}\chi^C,\qquad \phi^C = \(\half - \alpha^C - 
\beta^C\)F - \sum_i F^i(t^i)q^C_i - \lambda Q_X.\label{defphic}\eeq

In evaluating \myref{defX} we will use the fact that the
expression\footnote{It was noted in~\cite{gl} that the expression
  \myref{zero}, which is in fact the $T$-dependent part of the chiral
  anomaly found in~\cite{ss}, vanishes.  The authors of~\cite{ss}
  attribute~\cite{MS} this to their approximation that neglects higher
  order corrections.  However if these corrections take the form
  $g^i(T^i,\T^i) \to g^i(T^i,\T^i) + \Del^i(T^i,\T^i)$, our results
  our unchanged.  Note that the functional form of $\Del^i$ is
  severely restricted by the fact that it has to be invariant under
  T-duality.}
\beq \epsilon^{\mu\nu\rho\sigma} g^i_{\mu\nu}
g^{i}_{\rho\sigma} = 0,\label{zero}\eeq
\noindent
vanishes identically, and the expressions
\bea X^{i j} &=& \epsilon^{\mu\nu\rho\sigma}\im F^i g^i_{\mu\nu}
g^{j\ne i}_{\rho\sigma} =
4\epsilon^{\mu\nu\rho\sigma}\im F^i\pp_\mu g^i_\nu
\pp_\rho g^j_\sigma = 4\pp_\rho\(\epsilon^{\mu\nu\rho\sigma}\im F^i
\pp_\mu g^i_\nu g^j_\sigma\),\nnn
X^i &=& \half\epsilon^{\mu\nu\rho\sigma}\im F^i g^i_{\mu\nu}
X_{\rho\sigma} =
4i\pp_\rho\(\epsilon^{\mu\nu\rho\sigma}\im F^i
\pp_\mu g^i_\nu\Gamma_\sigma\),\nnn
X^{i a} &=& \epsilon^{\mu\nu\rho\sigma}\im F^i g^i_{\mu\nu}
F^a_{\rho\sigma}
= 4\pp_\rho\(\epsilon^{\mu\nu\rho\sigma}\im F^i
\pp_\mu g^i_\nu A^a_\sigma\),\label{epsilons}\eea  
\noindent
are total derivatives, where $A_\mu^a$ is an Abelian gauge field, and
\beq g^i = -\ln(t^i + \t^i), \qquad
g^i_\mu = -{\pp_\mu t^i - \pp_\mu \t^{\ibar}\over t^i
+\t^{\ibar}},\qquad g^i_{\mu\nu} = \pp_\mu g^i_\nu - \pp_\nu g^i_\mu.
\label{gdefs}\eeq

The full K\"ahler potential for $\dY$, with no anomalous \ux, is given
in~\cite{bg,gl}; here it takes the form
\bea K(\dY) &=& e^{\dot G}\(\sum_A e^{- g^a - q^a\vx}|\dY_A|^2 
+ \sum_I e^{- 2g^i}|\dY_I|^2 + \sum_N|\dY_N|^2\) + \ldots,\nnn 
g^a &=& \sum_n q_n^a g^n, \qquad \dot G = \dot\alpha K + \dot\beta g,
\qquad \dot\alpha +\dot\beta = 1,\label{KdY}\eea
\noindent
where $\dY_{N=1,2,3}$ (and their counterparts $\dZ^N$) are gauge
singlet PV fields needed~\cite{pv} to make the K\"ahler potential and
superpotential terms for $\dZ,\dY$ fully invariant, and the ellipsis
represents terms that make no contribution to the expressions given
below.  Using the sum rules in \myref{qsums} and \myref{sums} we obtain:
\bea \Tr\deta\Gamma_\alpha^{\dY} &=& - \[(N + 2)\dot\beta - A_1\]\ga,
\qquad \Tr\deta T_X^{\dY} = \Tr T_X,
\nnn \Tr\deta\Gamma^{\dY}_\alpha\Gamma^{\dY}_\beta &=&
- 2\dot\alpha\[\dot\beta(N + 2) - A_1\]X_\alpha g_\beta - 
\[\dot\beta^2(N + 2) - \dot\beta A_1 + A_2\]
g_\beta g_\alpha\ddd - B_2\sum_n g^n_\alpha g^n_\beta\nnn
\Tr\deta\Gamma^{\dY}_\alpha T_a &=& \del_{a X}\Tr T_X^{\dY}\dot\Ga -
Q_{1a}\ga, \qquad \dot\Ga = \dot\alpha\Xa + \dot\beta\ga.\label{log3}\eea
\noindent 

Using \myref{epsilons} and \myref{qsums}, the part of $X^{\dY}$ that is independent
of gauge charges takes the form:
\bea X^{\dY}_\chi &\ni& \half\[(N + 2) -2A_1\]
F\dot G\cdot\widetilde{\dot G} 
- \(A_1 - 2A_2\)F \dot G\cdot\tg - A_3F g\cdot\tg\mmm  
+ \mbox{total derivative}, \qquad \dot G _{\mu\nu} 
= \dot\alpha X_{\mu\nu} + \dot\beta g_{\mu\nu}.
\label{cubic2}\eea
\noindent
The modular weights for the $\psi$ satisfy
\noindent 
\bea \sum_{m,n}g^{n}q_n^{P_m} &=& g q^P_\psi,\qquad
\sum_{P}\eta^P_\psi q_l^{P_k}q_n^{P_k}q_n^{P_k} = 0,\nnn
\sum_{l,m,n}g^m g^n q_m^{P_l}q_n^{P_l} &=& (q_\psi^P)^2\sum_n g^n g^n.
\label{qpsisums} \eea
\noindent
{Like $X_\chi^{\dY}$, $X_\chi^\psi$} depends only on
$F,g_{\mu\nu}$ and $X_{\mu\nu}$, and \myref{log3} and \myref{cubic2}
can be cancelled by some combination of the fields in
\myref{pvfields}, with the condition
\noindent
\beq \sum_P\eta_\psi^P(q_\psi^P)^2 = B_2.\label{bcond}\eeq
\noindent
The pure T-moduli anomaly is given by
\noindent
\beq \Del L_1(F g^2) = {F\over8\pi^2}\Tr\eta_\psi\(1 - 
2\bar\gamma_\psi\)q_\psi^2\Omega_g, \qquad \chiproj\Omega_g  = 
\sum_n\gc_n\ga^n.\label{l1Fg2}\eeq
\noindent
Consistency with string results~\cite{JL} requires
\noindent
\beq \Tr\eta_\psi\(1 - 2\bar\gamma_\psi\)q_\psi^2 = -8\pi^2b
\label{qpsi2cond}\eeq

Finally, we require
\beq \Del L_1(Q_X g^2) = - {2\Lambda\over8\pi^2}\Tr\eta\bar Q_X
\Omega_f = \half\Lambda\dx\Omega_g.\label{l1Qg2}\eeq
\noindent
\newline
Using \myref{qpsisums}, the condition \myref{l1Qg2} requires
\noindent
\beq \sum_P\eta_\psi^P\bar Q_\psi^P(q_\psi^P)^2 = - 4\pi^2\dx.
\label{Qq2cond}\eeq
\noindent
All other other contributions to $\Del L_1$ are required to vanish.
\vskip .2in

We conclude this section by noting that cancellation of divergences
linear in the \ua\, field strengths is much simpler than for the
$\mathbb{Z}_7$ case considered in~\cite{gl}, as outlined below.

The gauge charges for the FIQS (~\cite{fiqs}) model are
listed\footnote{We have made some corrections to the \ua\, charges
  given in \myref{fiqs}.}  in \myapp{charges}. The universality of the
anomaly term quadratic in Yang-Mills fields strengths is guaranteed by
the universality condition \myref{uxconds}, as discussed in
\mysec{anom}.  Since gauge transformations commute with modular
transformations, a set of chiral multiplets $\Phi^b$ that transform
according to a nontrivial irreducible representation $R$ of a
nonabelian gauge group factor $\G_a$ have the same modular weights
$q^R_n$ such that
\beq \sum_{b\in R}q^b_n(T_a)^b_b = q^R_n(\Tr T_a)_R = 0.\eeq  
\noindent
Therefore terms linear in Yang-Mills field strengths occur only for
Abelian gauge group factors.  We need to cancel the $\dY$-loop
contribution to logarithmic divergences
\noindent
\beq \(\Tr\eta\sum_n q_n\ga^n T_a\)_{\dY} = - \sum_{b,n}q_n^b Q_a^b
g^n_\alpha = - Q_{1a}g_\alpha,\label{xan}\eeq 
\noindent 
and, dropping terms proportional to the last expression in
\myref{epsilons}, the relevant $\dY$ contributions to linear divergences:
\bea X^{\dY}_\chi &\ni& 
\sum_{a,b,n}Q^b_a\tF^a\cdot\[g^{n}q_n^b\(F - 2\sum_m q^b_m F^m\)
+ 2q^b_n F^n\(\dot G - \half X\)\]\eee
\sum_a\tF^a\cdot\lbr\[g\(1 + 2\dot\beta\) + X\(2\dot\alpha - 1\)\]
Q_{1a}F - 2\sum_n g^n F^n Q_{2a}\rbr,\label{xmn}\eea
\noindent
where we used \myref{qsums}.
The last term in \myref{xmn} is cancelled by 
\noindent
\bea X_\chi^{\psi} &\ni& -2\sum_{a,P,l,m,n}\eta_\psi^P Q_a^P q^{P_l}_m
q^{P_l}_n F^m\tF^a\cdot g^n = - 2\sum_{a,P}\eta_\psi^P Q_a^P(q^P)^2
\tF^a\cdot\sum_n g^n F^n,\eea
\noindent
provided 
\beq \sum_{P}\eta_\psi^P Q_a^P(q^P)^2 = - Q_{2a}.\eeq
\noindent 
The remaining terms in \myref{xmn}, as well as \myref{xan} can be
cancelled by a combination of the fields in \myref{pvfields}.  For $a
= X$ there are additional terms proportional to $(\Tr\eta T_X)_{P V} =
- \Tr T_X$.

\section{The final anomaly in the FIQS model}
\setcounter{equation}{0}

In \myapp{conds} we show that is possible to cancel all the ultraviolet
divergences from the $\dY$ fields with a choice of the set 
\myref{pvfields} such that the fields with noninvariant masses have
the properties
\beq\Tr\eta(\ln\cM)^{n>1} = \Del\Tr\eta(\ln\cM)^{n>1} = 
\Tr\eta(\Del\ln\cM)(\f_\alpha)^{n>0} =0.\label{zeros}\eeq
\noindent
Then, including the results of~\cite{gl}, the anomaly due to the
variation of \myref{SFtotanom} takes the form
\bea \del\L_{\rm anom} &=& \int d^4\theta E\(b F - \half\dx\Lambda\)\Omega
+ \int d^4\theta E b F\Omega',\eea
\noindent
where
\bea
\Omega &=& \Omega_{\rm Y M} - \Omega_{\rm G B} + \Omega_g,\nnn
\Omega' &=& - {b_{\rm spin}\over48b}\(4G_{\dot\beta\alpha}G^{\alpha\dot\beta} 
- 16R\bR + \D^2R + \bD^2\bR\) - {1\over8\pi^2 b}\Omega_D,\label{finalan}\eea
\noindent
where $\Omega_g$ is defined in \myref{l1Fg2}, and $b_{\rm spin}$ governs
the contributions from PV masses, as opposed to those arising from
uncancelled divergences:
\beq 8\pi^2 b_{\rm spin} = 8\pi^2 b +1,\eeq
\noindent
with $8\pi^2 b = 6$ in the FIQS model.  In the absence of an anomamous
\uo, $\Lambda =0$, the anomaly can be cancelled by the four
dimensional GS mechanism as described in~\cite{gl}.  However with $\Lambda\ne0$, the anomaly
as written in \myref{finalan} is no longer universal and cannot be
cancelled by the GS term alone.  However all of the ``D-terms'', in
other words the full expression $\Omega'$, can be removed~\cite{DB} by
adding counterterms to the Lagrangian, giving a universal anomaly
which can now be cancelled by the GS term.\footnote{The elimination of
  $\Omega_D$ further obviates the need for a modification of the
  linear-chiral duality transformation, a possibility condsidered in
  Appendix B of~\cite{gl} and Appendix E of~\cite{bg}.}

The results for the Gauss-Bonnet and Yang-Mills terms are 
well-established~\cite{prev} and result from the universality conditions
\myref{uxconds}.

\section{Conclusions}
\setcounter{equation}{0}

We have shown that a suitable choice of Pauli-Villars regulator fields
allows for a full cancellation of the chiral and conformal anomalies
associated, respectively, with the linear and logarithmic divergences
in the effective supergravity theory from a $\mathbb{Z}_3$ orbifold
compactification with Wilson lines and an anomalous \uo.

A future work~\cite{JL} will compare this result with that obtained directly
from string theory.

\vskip .3in
\noindent{\bf Acknowledgments.}  This work was supported in
part by the Director, Office of Science, Office of High Energy and
Nuclear Physics, Division of High Energy Physics, of the
U.S. Department of Energy under Contract DE-AC02-05CH11231, in part by
the National Science Foundation under grant PHY-1316783, and in part
by the European Union’s Horizon 2020 research and innovation program
under the Marie Skłodowska-Curie grant agreements No 690575 and No
674896.

\newpage
\appendix

\def\ksubsection{\Alph{subsection}}
\def\theequation{\ksubsection.\arabic{equation}} 

     
\catcode`\@=11

\def\thesubsection{\Alph{subsection}}
\def\thesubsubsection{\Alph{subsection}.\arabic{subsubsection}}
\noindent{\large \bf Appendix}

\setcounter{equation}{0}
\subsection{Conditions for the cancellation of ultraviolet divergences 
and the evaluation of the anomaly}\label{conds} 
\subsubsection{Notation}
	We pair PV fields according to their mass terms. A pair of PV fields ($\Phi^P$, $\Phi^{\prime P}$) has a superpotential coupling
		\bea
				W_{PV} =\sum_P \mu_P \Phi^{\prime P }\Phi^P
		\eea
	and a K\"ahler potential 
		\bea
			K_{PV} = \sum_P e^{f^P}\Phi^{*P}\Phi^P + \sum_P e^{f^{\prime P}}\Phi^{\prime *P}\Phi^{\prime P}, 
		\eea
	where
		\bea
				f^P = \alpha^P K + \beta^P g + \sum_n q^P_n g^n
		\eea
	with an identical definition holding for $f^{\prime P}$ but with primes on the constants $\{\alpha^P, \beta^P, q_n^P\}$.
	While we will not use it often, summing over the index $C$ means summing over PV fields and then their primed partners whereas summing over $P$ means summing over only the unprimed or primed fields, depending on the quantity being summed. For example,
	\bea
			\sum _C \eta^C \alpha^C = \sum_P \eta_P \alpha^P + \sum_P \eta_P \alpha^{\prime P}.
	\eea
\noindent
However, to reduce clutter, we will abbreviate the above. When summing over primed and unprimed fields, we will use ``$\Tr$". When summing over only primed or unprimed ones, we will use ``${\rm Sum}$". Thus the above would be written as
	\bea
			\text{Tr}\lbrack\eta \alpha\rbrack = \text{Sum}\lbrack\eta \alpha\rbrack + \text{Sum}\lbrack\eta \alpha^\prime\rbrack.
	\eea
	We will also encounter sums over various combinations of U(1) charges, U(1)$_X$ charges, and modular weights. To abbreviate these, especially when dealing with the quantum numbers of the light fields, we will define
	\bea
			Q_{1a} &=& \text{Sum}\lbrack\eta Q_a q_n\rbrack\\
			Q_{2a} + P_{2a}\delta_{nm} &=& \text{Sum}\lbrack\eta Q_a q_n q_m\rbrack\\
			R_a &=& \text{Sum}\lbrack\eta Q_a Q_X q_n\rbrack\\
			R_{ab} &=& \text{Sum}\lbrack\eta Q_a Q_b q_n\rbrack\\
			S_{a} &=&\text{Sum}\lbrack\eta Q_a Q_X\rbrack\\
			S_{ab} &=& \text{Sum}\lbrack\eta Q_a Q_b \rbrack.
	\eea

\subsubsection{Conditions for Regularization}
The terms we must cancel come from linear, logarithmic, and quadratic
divergences. It is helpful to organize these terms by forming subsets
based on whether terms depend on nonabelian gauge interactions,
nonanomalous Abelian gauge interactions, anomalous Abelian gauge
interactions, or none of the above. We will refer to these groupings
as nonabelian divergences, U(1)$_a$ divergences, U(1)$_X$
divergences, and modular divergences, respectively.  As an overview,
the divergences come from the terms
\bea
	&\text{Tr}\lbrack\eta\Gamma_\alpha\rbrack\\
	&\text{Tr}\lbrack\eta\Gamma_\alpha\Gamma_\beta\rbrack\\
	&\text{Tr}\lbrack\eta\Gamma_\alpha T_a\rbrack\\
	& \text{Tr}\lbrack\eta T_a T_b\rbrack\\
	& \text{Tr}\lbrack\eta Q_a\rbrack,
\eea
where
\bea
		\Gamma_{D \alpha}^C &=& -\frac{1}{8} \left(\bar{\mathcal{D}}^2 - 8R\right)\mathcal{D}_\alpha Z^i \Gamma^C_{Di}\\
		\phi^C &=& \left(\frac{1}{2} - \alpha^C - \beta^C\right) F - \sum_i F^i q_i^C -q^C_X \Lambda\\
		G_{\mu\nu} &=& \Gamma^C_{C\mu\nu} - \frac{1}{2}X_{\mu\nu}\delta^C_D - iF^a_{\mu\nu}(T_a)^C_D- iF^X_{\mu\nu}(Q_X)^C_D.
\eea
for our PV fields defined above.

The PV fields involved in this procedure are numerous. We take all of
the PV fields described in sections 3 and 4 of~\cite{bg} and supplement them
with further fields. However, to satisfy the divergences above, we
need only focus on the $\dot{Y}$ and $\hat{\phi}$ fields of~\cite{bg}. We now
group all the terms in the above expressions with our organizational
scheme.

\subsubsubsection{Modular Divergences\newline}

\noindent
 To cancel all the modular divergences, we require
	\bea
		 0 &=&  -\text{Tr}\bigg\lbrack\eta \beta\left(\frac{1}{2}-\alpha \right)^2 \bigg\rbrack -\text{Tr}\bigg\lbrack\eta q_n
		 	\left(\frac{1}{2}-\alpha \right)^2 \bigg\rbrack \label{L31}\\
		0 &=&  -\frac{1}{2} \text{Tr}\bigg\lbrack\eta(1-2 \alpha ) \beta  (1-2 \gamma )\bigg\rbrack
		+ \text{Tr}\bigg\lbrack\eta \beta  q_n(1-2 \alpha ) \bigg\rbrack
	\ddd	 -\frac{1}{2} \text{Tr}\bigg\lbrack\eta (1-2 \alpha ) (1-2 \gamma ) q_n\bigg\rbrack+\text{Tr}\bigg\lbrack \eta q_n q_m(1-2 \alpha ) \bigg\rbrack\label{L32}\\
		 0 &=&  \frac{1}{2}\text{Tr}\bigg\lbrack\eta \beta ^2 (1-2 \gamma )\bigg\rbrack
			 -\text{Tr}\bigg\lbrack\eta \beta ^2 q_n\bigg\rbrack
			 +\text{Tr}\bigg\lbrack\eta\beta  (1-2 \gamma ) q_n\bigg\rbrack
			 -2\text{Tr}\bigg\lbrack\eta  \beta  q_n q_m\bigg\rbrack 
\ddd +\frac{1}{2}\text{Tr}\bigg\lbrack\eta  (1-2 \gamma ) q_n q_m\bigg\rbrack
	 -\text{Tr}\bigg\lbrack\eta q_nq_mq_k\bigg\rbrack.\label{L33}
	\eea

\subsubsubsection{U(1)$_X$ Divergences\newline}

\noindent
 To cancel all the U(1)$_X$ Divergences, we need
\bea
		0 &=& \text{Tr}\lbrack\eta Q_X\rbrack\\
		0 &=& \text{Tr}\lbrack\eta Q_X\beta\rbrack + \text{Tr} \lbrack\eta Q_X q_m\rbrack\label{L41}\\
		0 &=& \text{Tr} \lbrack\eta Q_X\alpha\rbrack \label{L42}\\
		0 &=& \text{Tr}\left(\eta Q_X \left(\alpha - \frac{1}{2}\right)^2 \right)\label{L43}\\
		0 &=& -\text{Tr}\left(\eta Q_X\beta \left(\alpha - \frac{1}{2}\right)\right) +  \text{Tr}\left(\eta Q_X q_n \left(\alpha-\frac{1}{2}\right)\right)\label{L44}\\
		0 &=& \text{Tr}\left(\eta Q_X\beta^2\right) +2\text{Tr}\left(\eta Q_X q_n\beta\right) +\text{Tr}\left(\eta Q_X q_n q_m\right)\label{L45}\\
		0 &=& \text{Tr}\left(\eta Q_X^3\right)\label{L46}\\
		0 &=& \text{Tr}\left(\eta Q_X^2 \left(\frac{1}{2} - \gamma\right)\right) - \text{Tr}\left(\eta Q_X^2 q_n\right)\label{L47}\\
		0 &=& \text{Tr}\left(\eta Q_X^2 \left(\alpha - \frac{1}{2}\right)\right)\label{L48}\\
		0 &=& 2\text{Tr}\left(\eta Q_X \left(\alpha - \frac{1}{2}\right)\left(\frac{1}{2} - \gamma\right)\right) -2\text{Tr}\left(\eta Q_X q_n \left(\alpha - \frac{1}{2}\right)\right)\label{L49}\\
		0 &=& -2 \text{Tr}\left(\eta Q_X^2\beta\right) -2 \text{Tr}\left(\eta Q_X^2 q_n\right)\label{L410}\\
		0 &=&  2\text{Tr}\left(\eta Q_X \beta \left(\frac{1}{2} - \gamma\right)\right) -2 \text{Tr}\left(\eta Q_X q_n \beta\right)
			+2\text{Tr}\left(\eta Q_X q_n \left(\frac{1}{2} - \gamma\right)\right) \ddd -2 \text{Tr}\left(\eta Q_X q_n q_m\right).\label{L411}
\eea
\noindent
Note that only fields that have $\bar{Q}_X \neq 0$ will contribute to
Eq. \myref{L46}.

\subsubsubsection{Nonabelian Divergences\newline}

\noindent
 To cancel the nonabelian divergences, we need
\bea
		0 &=& \text{Tr}\lbrack\eta T_a T_b\rbrack \label{L51}\\
		0 &=& \text{Tr}\lbrack\eta Q_X T_aT_b\rbrack \label{L52}\\
		0 &=& \text{Tr}\bigg\lbrack\eta T_a T_b\left(\gamma - \frac{1}{2}\right)\bigg\rbrack, \label{L53}
\eea
where $T^a$ is a generator of a nonabelian gauge group factor.

\subsubsubsection{U(1)$_a$ Divergences\newline}

\noindent
\bea
	0 &=& \text{Tr}\lbrack\eta Q_a\rbrack\\
	0 &=& \text{Tr}\lbrack\eta Q_a \alpha\rbrack \label{L60}\\
	0 &=& \text{Tr}\lbrack\eta Q_a \beta\rbrack + \text{Tr}\lbrack\eta q_nQ_a\rbrack  \label{L61}\\
	0 &=& \text{Tr}\lbrack\eta Q_X Q_aQ_b\rbrack \label{L62}\\ 	
	0 &=&  \text{Tr}\lbrack\eta Q_X Q_a \beta\rbrack +  \text{Tr}\lbrack\eta Q_X q_n Q_a\rbrack \label{L63}\\
	0 &=& \text{Tr}\bigg\lbrack\eta Q_X Q_a \left(\alpha-\frac{1}{2}\right)\bigg\rbrack \label{L64}\\
	0 &=& -\text{Tr}\bigg\lbrack\eta Q_X Q_a \left(\frac{1}{2} - \gamma\right)\bigg\rbrack + \text{Tr}\lbrack\eta Q_X Q_a q_n\rbrack\label{L65}\\
	0 &=&  \text{Tr}\bigg\lbrack\eta Q_a \left(\alpha - \frac{1}{2}\right)\left(\left(\frac{1}{2} - \gamma\right)- q_n\right) \bigg\rbrack \label{L66}\\
	0 &=&  \text{Tr}\bigg\lbrack\eta Q_a \beta \left(\left(\frac{1}{2} - \gamma\right)-q_n\right)\bigg\rbrack +\text{Tr}\bigg\lbrack\eta Q_a q_n\left(\left(\frac{1}{2} - \gamma\right)-q_n\right)		\bigg\rbrack\label{L67}\\
	0 &=& \text{Tr}\bigg\lbrack\eta Q_a Q_b\left(\left(\gamma - \frac{1}{2}\right)+ q_n\right)\bigg\rbrack. \label{L68}
\eea
\noindent
In all of the above sets, we have assumed that the modular weights of
all PV fields satisfy sum rules reminiscent of those satisfied by the
light sector, \myref{qsums}.  Indeed, this will be baked directly into
our choice of PV fields. We have also used the total derivative
identities \myref{epsilons}.  In addition to the above conditions, we
must enforce the sum rules of~\cite{bg}:
\bea
		-N - 29 &=& \text{Tr}\lbrack\eta\rbrack\\
		-10 &=& \text{Tr}\lbrack\eta \alpha\rbrack\\
		-4 &=& \text{Tr}\lbrack\eta \alpha^2\rbrack\\
		0 &=& \text{Tr}\bigg\lbrack\eta \beta\bigg\rbrack\\
		0 &=& \text{Tr}\bigg\lbrack\eta \beta^2\bigg\rbrack\\
		0 &=& \text{Tr}\bigg\lbrack\eta \beta\alpha\bigg\rbrack.
\eea

\subsubsection{Conditions for Anomaly Matching}
By drawing an analogy with the calculation of~\cite{ss}, we infer that in 
four dimensions
the anomaly polynomial for the FIQS model has the form ~\cite{JL}
\bea I_{6} = \bigg(-{b\over4\pi}\sum_{i = 1}^3G_i + {\dx\over8\pi} F_X\bigg)\bigg(\text{tr}(R^2) &-& \sum_n( F^{SU(3)}_n)^2 - \sum_n (F^{SU(2)}_n)^2 -\sum_n (F^{SO(10)}_n)^2\nonumber\\
						&-& \sum_{a=1}^7 (F_a)^2 - (F_X)^2 + 2\sum_i G_i^2		\bigg)\eea
\noindent
where
\bea
	G_i &=& dZ_i\\
	Z_i &=&  \frac{1}{2i}\frac{d(T^i -\bar{T}^i)}{T^i +\bar{T}^i}\\	
\eea
and
\bea
	\text{tr}(R^2) &=& R^a_{\;\; b} R^b_{\;\; a} \\
			  &=& \frac{1}{4} R^\tau_{\;\; \epsilon\mu\nu}R^\epsilon_{\;\;\tau\rho\sigma}dx^\mu dx^\nu  dx^\rho  dx^\sigma \\
	(F_A)^2	 &=& \frac{1}{4} F_{A\mu\nu}F_{A\rho\sigma}dx^\mu dx^\nu dx^\rho dx^\sigma
			  	\eea
In the above, we have implicitly assumed wedge products in the multiplication of differential forms. To get the 4D anomaly from the 6-form anomaly polynomial, one goes through the usual descent equations:
\bea
		2\pi I_6 &=& dI_5\\
		\delta I_5 &=& dI_4
\eea			
For example, under a modular transformation, $Z_i \rightarrow Z_i + d\text{Im}(F^i)$ so that the modular-gravity-gravity anomaly has the form
\bea
	\int I_4 \supset \int -\frac{3}{32\pi^2}\bigg(\sum_{i=1}^3 \text{Im}(F^i)\bigg) R^\tau_{\;\; \omega\mu\nu}R^\omega_{\;\;\tau\rho\sigma} \epsilon^{\mu\nu\rho\sigma}\sqrt{g} d^4x
\eea
which is precisely what one would expect if one considers the modular-gravity-gravity anomaly to have the same form as a U(1)-gravity-gravity anomaly.
To match this anomaly, we look at the anomalous
contributions of PV fields with masses that are noninvariant under
modular and U(1)$_X$ transformations.  The general form of their
contribution is
\bea \mathcal{L}_{\rm anom} = \int d^4 \theta E (L_0 + L_1 +
L_r) \label{anomeq} \eea 
\noindent
with
\bea L_0 &=& \frac{1}{8\pi^2} \left(
\text{Tr}\lbrack\eta\ln(\mathcal{M}^2)\rbrack\Omega_0 + K
(\Omega_{GB}+\Omega_D) \right)\\ L_r &=& -\frac{1}{192\pi^2}
\text{Tr}\bigg\lbrack\eta\int d \ln(\mathcal{M})\Omega_r\bigg\rbrack.
\eea
\noindent 
Focusing on the second term of Eq. \myref{anomeq} , we again break
up terms based on whether they contribute to the U(1)$_X$ related anomalies
or the pure modular anomaly.

\subsubsubsection{U(1)$_X$ Anomaly Conditions\newline}

\noindent 
 To match the anomalies involving U(1)$_X$, we require
\bea
				0 &=& \frac{2}{3}\text{Tr}\bigg\lbrack\eta \bar{Q}_X \bigg(2\bar{\alpha}^2 + \bar{\alpha} - 
					3\alpha^2\bigg)\bigg\rbrack \label{L11}\\			
				0 &=& \frac{2}{3} \text{Tr}\bigg\lbrack\eta \bar{Q }_X \bigg(\bar{\beta} + 4\bar{\alpha}\bar{\beta} 
					-6\alpha \beta\bigg)\bigg\rbrack \label{L12}\\
				0 &=& \frac{2}{3}\text{Tr}\bigg\lbrack\eta\bar{Q }_X \left(2 \bar{\beta }^2 -3 \beta ^2 \right)\bigg\rbrack \label{L13}\\	
				0 &=& -4\text{Tr}\bigg\lbrack\eta\left( \alpha  \bar{Q }_X q_n\right)\bigg\rbrack \label{L14}\\  
				0 &=&-4 \text{Tr}\bigg\lbrack\eta\left( \beta  \bar{Q }_X q_n\right)\bigg\rbrack \\
8\pi^2\dx\del_{m n} &=& -2\text{Tr}\bigg\lbrack\eta\bigg(\bar{Q}_Xq_nq_m\bigg)\bigg\rbrack \label{L15}\\	
				0 &=& \frac{1}{3}\text{Tr}\bigg\lbrack\eta\bigg( Q_X  \left(-4 \bar{\alpha }+6 \alpha -1\right) \left(1-2 
					\bar{\gamma }\right)\bigg)\bigg\rbrack \label{L16}\\
				0 &=&\frac{2}{3} \text{Tr}\bigg\lbrack\eta  Q_X  \left(1-2 \bar{\gamma }\right) \left(3 \beta  Q_X -2 \bar{\beta } \bar{Q }_X\right))\bigg\rbrack \label{L17}\\
				0 &=& 2\text{Tr}\bigg\lbrack\eta\left(Q_Xq_n\left(1-2\bar{\gamma}\right)\right)\bigg\rbrack\label{L18}\\	
8\pi^2b &=& \frac{1}{3}\text{Tr}\bigg\lbrack\eta \left(1-2 \bar{\gamma }\right) \left(3 Q_X ^2-2 \bar{Q }_X^2\right)\bigg\rbrack\label{L19}\\
				 0 &=& \frac{2}{3}\text{Tr}\bigg\lbrack\eta \bar{Q }_X \left(4 \bar{\alpha } \bar{Q }_X+\bar{Q }_X-6 \alpha  Q_X \right)\bigg\rbrack		\label{L110}\\
			 	0 &=& \frac{1}{3}\text{Tr}\bigg\lbrack\eta \left(8 \bar{\beta } \bar{Q }_X^2-12 \beta  Q_X  \bar{Q }_X\right)\bigg\rbrack\label{L111}\\
				0 &=& -4\text{Tr}\bigg\lbrack\eta\left(Q_X \bar{Q}_Xq_n\right)\bigg\rbrack\\	
  -4\pi^2\dx &=&  \text{Tr}\bigg\lbrack\eta \left(\frac{4 \bar{Q }_X^3}{3}-2 Q_X ^2 \bar{Q }_X\right)\bigg\rbrack = 
				-\frac{2}{3}\text{Tr}\bigg\lbrack\eta Q_X^3\bigg\rbrack.\label{L112}
	\eea
\noindent
Note that the last term is fixed by cancellation of the linear
divergence term Eq.\myref{L46}.


\subsubsubsection{Pure Modular Anomaly Conditions\newline}
\noindent
 To match the pure modular anomaly, we require
\bea
	0 &=& \frac{1}{3}\text{Tr}\bigg\lbrack\eta \left(1-2 \bar{\gamma }\right)\left(-2 \bar{\alpha }^2-\bar{\alpha }+3 \alpha ^2\right) \bigg\rbrack \label{L21}\\
	0 &=& \frac{1}{3}\text{Tr}\bigg\lbrack\eta\left(1-2 \bar{\gamma }\right) \left(3 \beta ^2-2 \bar{\beta }						^2\right) \bigg\rbrack + 2\text{Tr}\bigg\lbrack\eta  \beta  \left(1-2 \bar{\gamma }\right) q_n\bigg\rbrack  \label{L22}\\
	0 &=& \frac{1}{3}\text{Tr}\bigg\lbrack\eta\left(1-2 \bar{\gamma }\right) \left(6 \alpha  \beta -\left(4 \bar{\alpha }+1\right) \bar{\beta }\right)\bigg\rbrack + 2\text{Tr}\bigg\lbrack\eta \alpha  \left(1-2 \bar{\gamma }\right) q_n\bigg\rbrack \label{L23}\\
-8\pi^2b\del_{m n}  &=& \text{Tr}\bigg\lbrack\eta q_m q_n\left(1-2 \bar{\gamma }\right) \bigg\rbrack. \label{L24}
\eea
\noindent
As for the third term of Eq. \myref{anomeq}, we need it to vanish
identically. This can be achieved so long as the following are
satisfied
\bea
		0 &=&  \text{Tr}\bigg\lbrack\eta x (1-2\bar{\gamma})^2\bigg\rbrack \label{M1}\\
			0 &=&  \text{Tr}\bigg\lbrack\eta x \bar{q}_X(1-2\bar{\gamma})\bigg\rbrack\label{M2a}\\
			0 &=&  \text{Tr}\bigg\lbrack\eta x \bar{q}_X^2\bigg\rbrack\label{M3}\\
			0 &=&\text{Tr}\bigg\lbrack\eta \bar{\alpha}\bar{\beta}(1-2\bar{\gamma})\bigg\rbrack\label{M4}\\	
			0 &=&\text{Tr}\bigg\lbrack\eta \bar{\alpha}\bar{\beta}\bar{q}_X)\bigg\rbrack\label{M5}\\
			0 &=& \text{Tr}\bigg\lbrack\eta\bar{\beta}^k (1-2\bar{\gamma})\bigg\rbrack\label{M6}\\
			0 &=& \text{Tr}\bigg\lbrack\eta \bar{\beta}^k\bar{q}_X \bigg\rbrack\label{M7}
                        \text{Tr}\bigg\lbrack\eta\bar{\beta}^3\bar{q}_X\bigg\rbrack,
                        \eea 
\noindent 
where $x= 1, \bar{\alpha},
                        \bar{\beta}, \bar{q}_X, \bar{\alpha}^2,
                        \bar{\beta}^2, \bar{q}_X^2,
                        \bar{\alpha}\bar{\beta},\bar{\alpha}\bar{q}_X,
                        \bar{\beta}\bar{q}_X$ and $k = 1,2,3$.
 

\setcounter{equation}{0}
\subsection{Solution to the Pauli-Villars Regularization Conditions}
\label{sols}
We will now elucidate a solution to the system described above. The solution consists of sets  $S_a$, $a= 1,2,\ldots$ of PV fields that address each of the divergence and anomaly sets of conditions more or less separately. For example, it is possible to introduce PV fields that cancel only the nonabelian divergences and contribute to no other conditions. We will try to follow the same strategy for all the sets of conditions described above. It is not entirely possible to do so - for example, fields that solve the modular anomaly conditions will generically contribute to modular divergences. Of course, this is far from the only way to tackle the system, but it is straightforward method to illustrate that a solution can be found. To this end, we define the notion of clone fields for PV fields. For a given pair of PV fields $\left(\Phi^P , \Phi^{\prime P}\right)$, we define clone fields $\left(\Phi^P_{cl} , \Phi^{\prime P}_{cl}\right)$ that have almost the same parameters ($\alpha$, $\beta$, $q_n$, $\ldots$) and quantum numbers as the original pair but with negative signature. We say almost here because this notion is only useful if the $\left(\Phi^P , \Phi^{\prime P}\right)$ have quantum numbers different from the clones so that the two sets cancel each other's contributions to some subset of the conditions, but not all conditions. As a concrete example, which will be described below, one can introduce PV fields with nonabelian gauge interactions to eliminate divergences associated with those same interactions. One can then introduce clone PV fields without gauge interactions that exactly cancel the contributions of the gauge charged PV fields to all other terms. The primary advantage of this technique is tidiness.

\subsubsection{PV Fields for U(1)$_X$ Anomaly Matching}
The fields described here will satisfy Eqs. \myref{L11}--\myref{L112} and
will contribute to some of the U(1)$_X$ divergence conditions
\myref{L41}--\myref{L411}. In particular, only PV fields with
$\bar{Q}_X \neq 0 $ contribute to Eq. \myref{L46}, so this condition will
be taken care of by this sector only. The sets of PV fields we need are
\begin{itemize}
	\item $S_1$: A set of PV fields with 
          modular invariant masses,$\alpha_1 = \alpha_1^\prime =
          \bar{\gamma}_1 = 1/2$, and $\bar{q}^{(1)}_n = 0$ and modular
          weights of the form $(q^{(1)})^C_m = q^P_{(1)} \delta^n_m$
          and clone fields with no U(1)$_X$.
	\item $S_2 $ : A set of PV fields with $\bar{\alpha}_2 =
          \bar{\beta}_2 = \bar{\gamma}_2 = \bar{Q}^{(2)}_X =
          (q^{(2)})^C_n = 0$ and clone fields with no U(1)$_X$ charge.
\end{itemize}
 We then place the following conditions on the parameters for these fields:
\bea
{\rm Sum}\[Q_X^{(L)}\] &=& - {\rm Sum}\[\eta Q_X^{(1)}\] \\
\text{Sum}\bigg\lbrack(Q_X^L)^3\bigg\rbrack &=&
-\text{Tr}\bigg\lbrack\eta_1 (Q_X^{(1)})^3\bigg\rbrack\\
0 &=& \Tr\[\eta_1\bar{Q}_X^{(1)}\(1 - 3\alpha_1^2\)\]\\
0 &=& \[\eta_1 \bar{Q}_X^{(1)}\alpha_1 q_n^{(1)}\] \\
		0 &=& \text{Tr}\bigg\lbrack\eta_1\alpha_1 \bar{Q}_X^{(1)}
 {Q}_X^{(1)}\bigg\rbrack
\\
		0 &=& \text{Tr}\bigg\lbrack\eta (\bar{Q}_X^{(1)})^2\bigg\rbrack\\
		0 &=& \text{Tr}\bigg\lbrack\eta (\bar{Q}_X^{(1)})^3\bigg\rbrack\\
		0 &=& \text{Tr}\bigg\lbrack\eta (\bar{Q}_X^{(1)})^4\bigg\rbrack\\
		0 &=& \text{Tr}\bigg\lbrack\eta \bar{Q}^{(1)}_X q_n^{(1)}\bigg\rbrack\\
		0 &=& \text{Tr}\bigg\lbrack\eta \bar{Q}^{(1)}_XQ^{(1)}_X q^{(1)}_n\bigg\rbrack\\
- 4\pi^2\dx\del_{n m} &=& 
\text{Tr}\bigg\lbrack\eta \bar{Q}^{(1)}_Xq_n^{(1)}q_m^{(1)}\bigg\rbrack\\
2\pi^2\dx &=& 
-\frac{1}{3}\text{Sum}\bigg\lbrack(Q_X^{(L)})^3\bigg\rbrack= \text{Tr}\bigg\lbrack\eta \bar{Q}^{(1)}_X (Q_X^{(1)})^2\bigg\rbrack.
\eea
\noindent
Once again, the first condition is a linear divergent term that can only be cancelled by fields with masses that are noninvariant under U(1)$_X$. This in turn forces the correct coefficient for the pure U(1)$_X$ anomaly in the last condition. While the second set must satisfy
\bea
	0 &=& \text{Tr}\lbrack\eta_2\rbrack	\\
	0 &=& \text{Tr}\bigg\lbrack\eta_2\alpha_2 Q_{X}^{(2)}\bigg\rbrack	\\
	0 &=& \text{Tr}\bigg\lbrack\eta_2\beta_2 Q_X^{(2)}\bigg\rbrack	\\
  8\pi^2b &=& \text{Tr}\bigg\lbrack\eta_2(Q_X^{(2)})^2\bigg\rbrack.	
\eea
\noindent
The first condition here comes from Eq. \myref{M1} and potentially can
be relaxed.

\subsubsection{PV Fields for Modular Anomaly Matching}
The fields described here will satisfy conditions
\myref{L21}--\myref{L24} and contribute to the modular divergence
conditions \myref{L31}--\myref{L33}. The sets are
\begin{itemize}
	\item $S_3$: A set of pairs of PV fields with $\beta_3 =
          \beta_3^\prime = 0$, $q_n^{(3)} = q^{\prime (3) }_n = 0$.
	\item $S_4$: A set of pairs of PV fields with $\alpha_4 =
          \alpha_4^\prime = \beta_4 = \beta^\prime_4 = \bar{q}^n_4 =
          0$, $(q^{(4)})^C_m = (q^{(4)})^P\delta^n_m$ , and clone
          fields with no modular weights.
\end{itemize}
These fields will also contribute to the modular divergence conditions, as outlined below. We also have to consider the $\hat{\phi}$ fields of~\cite{bg} here
since they have noninvariant masses under a modular
transformation. These fields have no $\beta$ or modular weight
parameters but do have $f_{\hat{\phi}} = \hat{\alpha}K$. then the
conditions the $S_3$, $S_4$, and $\hat{\phi}$ fields must satisfy are
\bea
	0 &=& \text{Tr}\bigg\lbrack\hat{\eta}(1-2\bar{\hat{\alpha}})^2\bigg\rbrack + \text{Tr}\bigg\lbrack\eta_3 (1-2\bar{\alpha}_3)^2\bigg\rbrack\\
	0 &=& \text{Tr}\bigg\lbrack\hat{\eta}\bar{\hat{\alpha}}(1-2\bar{\hat{\alpha}})^2\bigg\rbrack + \text{Tr}\bigg\lbrack\eta_3\bar{\alpha}_3 (1-2\bar{\alpha}_3)^2\bigg\rbrack\\
	0 &=& \text{Tr}\bigg\lbrack\hat{\eta}\bar{\hat{\alpha}}^2(1-2\bar{\hat{\alpha}})^2\bigg\rbrack + \text{Tr}\bigg\lbrack\eta_3\bar{\alpha}_3^2 (1-2\bar{\alpha}_3)^2\bigg\rbrack\\
	0 &=&  \text{Tr}\bigg\lbrack\hat{\eta} \left(1-2 \bar{\hat{\alpha} }\right)\left(-2 \bar{\hat{\alpha} }^2-\bar{\hat{\alpha} }+3 \hat{\alpha} ^2\right) \bigg\rbrack + \text{Tr}\bigg\lbrack\eta_3 \left(1-2 \bar{\alpha }_3\right)\left(-2 \bar{\alpha }_3^2-\bar{\alpha }_3+3 \alpha_3 ^2\right) \bigg\rbrack
\eea
 and
 \bea
 		-8\pi^2b &=& \text{Tr}\bigg\lbrack\eta_4 q^P_4 q^P_4\bigg\rbrack =  
                2\text{Sum}\bigg\lbrack\eta_4 q^P_4 q^P_4\bigg\rbrack.
 \eea

\subsubsection{PV Fields for the Regulation of Modular Divergences}
Here we introduce fields that can cancel the contributions to
Eqs. \myref{L31}--\myref{L33} from the $\dot{Y}$, $S_3$, and $S_4$ and
contribute to the sum rules in Eqs.  (3.37), (3.38) and (A.16)
of~\cite{bg}. The only new set we introduce here is
 \begin{itemize}
 	\item $S_5$ : A set of pairs of PV fields with $\bar{\gamma}_5
          = \frac{1}{2}$ and $(\bar{q}^{(5)})^C_n = 0$ with
          $(q^{(5)})^C_m = (q^{(5)})^P \delta^n_m$.
 \end{itemize}
 Then the conditions we must satisfy are
{\footnotesize
\bea
	0 &=& (N+2)\dot{\beta} \left(\frac{1}{2} - \dot{\beta}\right)^2 - A_1 \left(\frac{1}{2} - \dot{\beta}\right)^2 - \text{Sum}			\bigg\lbrack\eta_5 \beta_5 \left(\frac{1}{2}- \alpha_5\right)^2\bigg\rbrack - \text{Sum}\bigg\lbrack\eta_5\beta^			\prime_5 \left(\frac{1}{2} - \alpha^\prime_5\right)^2\bigg\rbrack\nonumber\\ 
	&& - \text{Sum}\bigg\lbrack\eta_5q^P_5 \left(\frac{1}{2} - \alpha_5\right)^2\bigg\rbrack + \text{Sum}\bigg\lbrack\eta_5 			q^P_5\left(\frac{1}{2} - \alpha^\prime_5\right)^2\bigg\rbrack\\
	0 &=& (N+2)\dot{\beta}\left(\frac{1}{2} - \dot{\beta}\right) - A_1 \left(\frac{1}{2} - \dot{\beta}\right) - 2A_2\dot{\beta}\left(\frac{1}{2}-\dot{\beta}\right) + 2A_2\left(\frac{1}{2} - \dot{\beta}\right)\nonumber\\
		&&-\frac{1}{2}\left(\text{Sum}\bigg\lbrack\eta_5 \beta_5(1-2\alpha_5)(1-2\gamma_5)\bigg\rbrack + \text{Sum}\bigg\lbrack\eta_5 \beta_5^\prime (1-2\alpha_5^\prime)(1-2\gamma^\prime_5)\bigg\rbrack\right) \nonumber\\
		&& +\text{Sum}\bigg\lbrack\eta_5 q^P_5 \beta_5 (1-2\alpha_5)\bigg\rbrack -\text{Sum}\bigg\lbrack\eta_5 q^P_5 \beta^\prime_5 (1-2\alpha^\prime_5)\bigg\rbrack
- \frac{1}{2}\text{Sum}\bigg\lbrack\eta_5 q^P_5(1-2\alpha_5)(1-2\gamma_5)\bigg\rbrack
\ddd  - \half\text{Sum}\bigg\lbrack\eta_5 q^P_5 (1-2\alpha_5^\prime)(1-2\gamma^\prime_5)\bigg\rbrack
		 + \text{Sum}\bigg\lbrack\eta_5 q^P_5 q^P_5 (1-2\alpha_5)\bigg\rbrack + \text{Sum}\bigg\lbrack\eta_5 q^P_5 q^P_5 (1-2\alpha^\prime_5)\bigg\rbrack
\ddd + 2\text{Sum}\bigg\lbrack\eta_4 q^P_4 q^P_4\bigg\rbrack\\
	0 &=& (N+2)\frac{\dot{\beta}^2}{2} - A_1\dot{\beta} + \frac{A_2}{2}  - A_1\dot{\beta}^2 + 2A_2 \dot{\beta} - A_3 + \frac{1}{2} \left( \text{Sum}\bigg\lbrack\eta_5 \beta^2_5 (1-2\gamma_5)\bigg\rbrack + \text{Sum}\bigg\lbrack\eta_5 \beta^{\prime 2}_5 (1-2\gamma^\prime_5)\bigg\rbrack\right)\nonumber\\
		&& - \left(\text{Sum}\bigg\lbrack\eta_5 q^P_5 \beta^2_5\bigg\rbrack - \text{Sum}\bigg\lbrack\eta_5 q^P_5 \beta^{\prime 2}_5\bigg\rbrack\right) + \left(\text{Sum}\bigg\lbrack\eta_5 \beta_5 q^P_5 (1-2\gamma_5)\bigg\rbrack - \text{Sum}\bigg\lbrack\eta_5 \beta^\prime_5 q^P_5 (1-2\gamma^\prime_5)\bigg\rbrack\right)\nonumber\\
		& &-2 \left(\text{Sum}\bigg\lbrack\eta_5 \beta_5 q^P_5 q^P_5\bigg\rbrack +\text{Sum}\bigg\lbrack\eta_5 \beta^\prime_5 q^P_5 q^P_5\bigg\rbrack\right)  + \frac{1}{2} \left(\text{Sum}\bigg\lbrack\eta_5 (1-2\gamma_5)q^P_5 q^P_5\bigg\rbrack + \text{Sum}\bigg\lbrack\eta_5 (1-2\gamma^\prime_5)q^P_5 q^P_5\bigg\rbrack\right)\ddd + \text{Sum}\bigg\lbrack\eta_4 q^P_4 q^P_4\bigg\rbrack.
\eea}
\noindent
 We include an explicit P in the modular weights simply to remind
 ourselves that we sum over the ``P" index and not the ``n" index
 since C = (P,n).
 
\subsubsection{PV Fields for the Regulation of U(1)$_X$ Divergences}
Here we introduce fields that cancel the contributions to
Eqs. \myref{L41}--\myref{L411} from the $\dot{Y}$, $S_1$, and $S_2$. Note
that we will omit Eq. \myref{L46} since has been taken care of above. We
introduce the following set:
\begin{itemize}
		\item $S_6$: A set of pairs of PV fields with
                  $Q^{(6)}_X = -Q_X^{\prime (6) }$ and
                  $\bar{q}^{(6)}_n = 0$ and clone fields without
                  U(1)$_X$ charge.
\end{itemize}
Then the conditions we must satisfy are
{\footnotesize
\bea
		0 &=& 12C^\prime_{GS}\dot{\beta} + 2\text{Sum}\bigg\lbrack\eta_2 Q^X_{(2)}\beta_2\bigg\rbrack + \text{Sum}\bigg\lbrack\eta_6 Q^X_{(6)}\beta_6\bigg\rbrack - \text{Sum}\bigg\lbrack\eta_6 Q^X_{(6)}\beta^\prime_6\bigg\rbrack\\
		0 &=& 12C^\prime_{GS}(1-\dot{\beta}) + 2\text{Sum}\bigg\lbrack\eta_2 Q^X_{(2)}\alpha_2\bigg\rbrack  + \text{Sum}\bigg\lbrack\eta_6 Q^X_{(6)}\alpha_6\bigg\rbrack - \text{Sum}\bigg\lbrack\eta_6 Q^X_{(6)}\alpha^\prime_6\bigg\rbrack\\
0 &=& 12C^\prime_{GS}\left(\frac{1}{2}-\dot{\beta}\right)^2 + \text{Sum}\bigg\lbrack\eta_6 Q^X_{(6)}\left(\alpha_6 - \frac{1}{2}\right)\bigg\rbrack - \text{Sum}\bigg\lbrack\eta_6 Q^X_{(6)}\left(\alpha^\prime_6 - \frac{1}{2}\right) \bigg\rbrack\\
	0 &=& 12C^\prime_{GS} \dot{\beta}\left(\frac{1}{2} - \dot{\beta}\right) + Q_{1X}^{(L)}\left(\frac{1}{2}-\dot{\beta}\right) + \text{Sum}\bigg\lbrack\eta_2 Q^X_{(2)}q^P_2\bigg\rbrack + \text{Sum}\bigg\lbrack\eta_6 Q^X_{(6)}\beta_6 \left(\alpha_6-\frac{1}{2}\right)\bigg\rbrack 
\ddd- \text{Sum}\bigg\lbrack\eta_6 Q^X_{(6)}\beta_6^\prime \left(\alpha_6^\prime-\frac{1}{2}\right)\bigg\rbrack - \text{Sum}\bigg\lbrack\eta_6 Q^X_{(6)}q^P_6 \left(\alpha_6 - \frac{1}{2}\right)\bigg\rbrack - \text{Sum}\bigg\lbrack\eta_6 Q^X_{(6)}q^P_6 \left(\alpha_6^\prime - \frac{1}{2}\right)\bigg\rbrack\\
	0 &=& 12\dot{\beta}^2 C^\prime_{GS}  - 2\dot{\beta} Q_{1X}^{(L)} + Q_{2X}^{(L)} + \text{Sum}\bigg\lbrack\eta_1 Q^X_{(1)}q^P_1 q^P_1\bigg\rbrack + \text{Sum}\bigg\lbrack\eta_1 Q^{\prime X}_{(1)}q^P_1 q^P_1\bigg\rbrack + \text{Sum}\bigg\lbrack\eta_6 Q^X_{(6)} \beta_6^2\bigg\rbrack 
\ddd - \text{Sum}\bigg\lbrack\eta_6 Q^X_{(6)} \beta_6^{\prime 2}\bigg\rbrack
		+ 2 \text{Sum}\bigg\lbrack\eta_6 Q^X_{(6)} \beta_6 q^P_6\bigg\rbrack + 2\text{Sum}\bigg\lbrack\eta_6 Q^X_{(6)} \beta^\prime_6 q^P_6\bigg\rbrack \\
	0 &=& \frac{1}{2} \text{Tr}\bigg\lbrack (Q^X_{(L)})^2\bigg\rbrack - R_X^{(L)} 	- \text{Sum}\bigg\lbrack\eta_1 (Q^X_{(1)})^2 q^P_1\bigg\rbrack	 + \text{Sum}\bigg\lbrack\eta_1 (Q^{\prime X}_{(1)})^2 q^P_1\bigg\rbrack + \text{Sum}\bigg\lbrack\eta_2 (Q^X_{(2)})^2 \bigg\rbrack 
\ddd + \text{Sum}\bigg\lbrack\eta_6 (Q^X_{(6)})^2 \left(\frac{1}{2} - \gamma_6\right)\bigg\rbrack
+ \text{Sum}\bigg\lbrack\eta_6 (Q^X_{(6)})^2 \left(\frac{1}{2} - \gamma_6^\prime\right)\bigg\rbrack\\
	0 &=& {\text{Tr}\bigg\lbrack (Q^X_{(L)})^2\left(\frac{1}{2} - \dot{\beta}\right)
\bigg\rbrack} - \text{Sum}\bigg\lbrack\eta_2 (Q^X_{(2)})^2\bigg\rbrack + \text{Sum}\bigg\lbrack\eta_6 (Q^X_{(6)})^2 \left(\alpha_6 - \frac{1}{2}\right) \bigg\rbrack\ddd 
+\text{Sum}\bigg\lbrack\eta_6 (Q^{\prime X}_{(6)})^2 \left(\alpha_6 - \frac{1}{2}\right) \bigg\rbrack\\
	0 &=& - \frac{1}{2} Q_{1X}^{(L)}\left(\frac{1}{2} - \dot{\beta}\right) + R_X^{(L)}\left(\frac{1}{2} - \dot{\beta}\right) + \text{Sum}\bigg\lbrack\eta_2 Q^X_{(2)}(\alpha_2 + \gamma_2)\bigg\rbrack + \text{Sum}\bigg\lbrack\eta_2 Q^X_{(2)}q^P_2\bigg\rbrack\nonumber\\
		& &+ \text{Sum}\bigg\lbrack\eta_6 Q^X_{(6)}\left(\alpha_6 - \frac{1}{2}\right)\left(\frac{1}{2} - \gamma_6\right)\bigg\rbrack - \text{Sum}\bigg\lbrack\eta_6 Q^X_{(6)}\left(\alpha_6^\prime - \frac{1}{2}\right)\left(\frac{1}{2} - \gamma_6^\prime\right)\bigg\rbrack
		 -\text{Sum}\bigg\lbrack\eta_6 Q^X_{(6)} q^P_6 \left(\alpha_6 - \frac{1}{2}\right)\bigg\rbrack\nonumber\\
		 & &- \text{Sum}\bigg\lbrack\eta_6 Q^X_{(6)} q^P_6 \left(\alpha_6^\prime - \frac{1}{2}\right)\bigg\rbrack\\
	0 &=&  - \dot{\beta} \text{Tr}\bigg\lbrack (Q^X_{(L)})^2\bigg\rbrack + R_X^{(L)} + \text{Sum}\bigg\lbrack\eta_1 (Q^X_{(1)})^2q^P_1\bigg\rbrack - \text{Sum}\bigg\lbrack\eta_1 (Q^{\prime X}_{(1)})^2q^P_1\bigg\rbrack + \text{Sum}\bigg\lbrack\eta_6 (Q^X_{(6)})^2\beta_6\bigg\rbrack 
\ddd + \text{Sum}\bigg\lbrack\eta_6 (Q^X_{(6)})^2\beta_6^\prime\bigg\rbrack \\
	0 &=& - 6\dot{\beta}C^\prime_{GS} + \beta Q_{1X}^{(L)} + \frac{1}{2} Q_{1X}^{(L)} - Q_{2X}^{(L)} - \text{Sum}\bigg\lbrack\eta_1 Q^X_{(1)}q^P_1 q^P_1\bigg\rbrack- \text{Sum}\bigg\lbrack\eta_1 Q^{\prime X}_{(1)}q^P_1 q^P_1\bigg\rbrack	
	+ \text{Sum}\bigg\lbrack\eta_2 Q^X_{(2)}\beta_2\bigg\rbrack\ddd
 + \text{Sum}\bigg\lbrack\eta_2 Q^X_{(2)}q^P_2\bigg\rbrack + \text{Sum}\bigg\lbrack\eta_6 Q^X_{(6)}\beta_6 \left(\frac{1}{2} - \gamma_6\right)\bigg\rbrack - \text{Sum}\bigg\lbrack\eta_6 Q^X_{(6)}\beta_6^\prime \left(\frac{1}{2} - \gamma_6^\prime\right)\bigg\rbrack - \text{Sum}\bigg\lbrack\eta_6 Q^X_{(6)}\beta_6 q^P_6\bigg\rbrack\nonumber\\
		& &- \text{Sum}\bigg\lbrack\eta_6 Q^X_{(6)}\beta_6^\prime q^P_6\bigg\rbrack + \text{Sum}\bigg\lbrack\eta_6 Q^X_{(6)} q^P_6\left(\frac{1}{2} - \gamma_6 \right)\bigg\rbrack + \text{Sum}\bigg\lbrack\eta_6 Q^X_{(6)} q^P_6\left(\frac{1}{2} - \gamma_6^\prime \right)\bigg\rbrack.
\eea}

\subsubsection{PV Fields for the Regulation of Nonabelian Divergences}
Here we introduce fields to cancel Eqs. \myref{L51}-\myref{L53}. We
introduce a separate PV set for each of the nonabelian factors of the
FIQS gauge group as follows
\begin{itemize}
		\item $S_7$: A set of pairs of PV fields in the
                  fundamental of SU(3) ( anti-fundamental for the
                  primed fields) with no modular weights, uniform
                  constants, and clone fields with no gauge charges.
                  By uniform coefficients, we mean that $\alpha^C$ and
                  $\beta^C$ are independent\ \ of index within the
                  set: $\alpha^C = \alpha$ and $\beta^C = \beta$.
		\item $S_8$: A set of pairs of PV fields in the
			fundamental of SU(2) with no modular weights,
			uniform constants, and clone fields with no
			gauge charges.  \item $S_9$: A set of pairs of
			PV fields in the $\bf{16}$ (and
			$\overline{\bf{16}}$ for primed fields) of SO(10)
			and a set of pairs of PV fields in the
			$\bf{10}$ of SO(10), all with no modular
			weights, uniform constants, and clone fields
			with no gauge charges.  \item $S_{10}$: A set
			of PV fields with $\gamma = \gamma^\prime =
			1/2$, zero modular weights, a nonzero trace
			U(1)$_X$ charge matrix, and charged under the
			nonabelian gauge groups in the same reps as
			the light fields and clone fields without
			nonabelian gauge charges.
\end{itemize}
Let us discuss this choice briefly.  First we need to check the
number of fields in a given representation. This is because we care
about the quantity
\bea
	C_{(\mathcal{G})}^M = C^m_{(\mathcal{G})}N_{(\mathcal{G})},
\eea
\noindent
which comes from the first term in the list above. The technique in~\cite{gl}
relies on having an even number of light fields in a given
representation for all the gauge factors. Let us check if this is the
case for the FIQS model. See Appendix C for a detailed breakdown of the FIQS spectrum.
  For the SU(3) of FIQS, the total number of
triplets charged under this gauge group is 
\bea
	N_{Q_L}^{SU(3)}+ N_{u_L}^{SU(3)}+ N_{u_2}^{SU(3)} +\sum_{i=1}^2  N_{d_i}^{SU(3)} + \sum_{j=1}^4 N_{D_j}^{SU(3)} + \sum_{j=1}^2 N_{\bar{D}_j}^{SU(3)}\nonumber\\
	\;\;\;= 6+3+12+15 = 36.
\eea
For the SU(2) of FIQS, there are
\bea
	N_{Q_L}^{SU(2)}+ \sum_{i=1}^4 N^{SU(2)}_{\bar{G}_i}+ \sum_{i=1}^5 N^{SU(2)}_{G_i}+ \sum_{i=1}^4 N^{SU(2)}_{F_i}\nonumber\\ \;\;\;=	9 + 3 + 33 + 3 = 48
\eea
\noindent
doublets. Note that we have used the fact that each state in the
table of Appendix C has a degeneracy of 3, with the exception of
the states $Y_1$, $Y_2$, and
$Y_3$. The number of states charged under the SU(3) and SU(2) groups
are indeed even, but this is not the case for SO(10), since there are
only 3 $\bf{16}$'s charged under this gauge factor.  To resolve this,
we first list the Casimirs for the first few SO(10) reps.
\bea
		\text{Fundamental } & {\bf{10}}: C_{10} = 1\\
		\text{Spinor } & {\bf{16}}: C_{16} = 2\\
		\text{Adjoint } & {\bf{45}}: C_{45} = 8 
\eea
\noindent
Note that these satisfy the sum rule (5.12) of~\cite{bg} when considering the
fields charged under SO(10): 
\bea C_{45} - 3C_{16} + 2 C_{16}\sum_i
\delta^i_n = 8 -6 + 4 = 6 \eea
\noindent
The first divergence we cancel is $\text{Tr}(\eta T_a
T_b)$. The $\dot{Y}$ give the negative of the contribution of the
light fields, so in the case of SO(10) this trace is simply $-3C_{16} = -6$. Since PV
fields come in pairs, we cancel this with at least 2 fields and we
have
\bea
		3C_{16} = 2\sum_P\eta^P C^P
\eea
\noindent
Thus, we have two options. We can have a PV pair in the $\bf{16}$ (and
$\overline{\bf{16}}$) plus a PV pair in the $\bf{10}$ or we can have 3
pairs of PV fields in the $\bf{10}$. The other divergence from gauge
interactions we have to get rid of is the linear divergence
proportional to the Casimir.  We note that the $\dot{Y}$'s here give
\bea
		(-1)\left(\frac{F}{2} -F + \sum_n q^a_n F^n\right)C_{(\mathcal{G}_a)} = \left(-\frac{1}{2}\right)(C_{GS} - C_{\mathcal{G}})
\eea
\noindent
since $\dot{\alpha} + \dot{\beta} =1$. The overall sign is the sum of
the signatures. Cancellation then requires
\bea
		\frac{C_{GS} - C_{\mathcal{G}}}{2} &=& \sum_C \eta^C C_{\mathcal{G}_C} \left(\frac{1}{2} -\gamma^C\right)\\
						&=& \sum_P \eta^P C_{\mathcal{G}_P} \left(1 -2\bar{\gamma}^P\right).
\eea
\noindent
provided that the PV fields have no modular weights. The first sum is over all PV
fields whereas the second is over PV pairs. Both of our potential
solutions can work here since we have either 1 or 2 free parameters in
the $\gamma$'s. In the list of sets of PV fields above, we opted for the combination of PV fields in the $\bf{10}$ and $\bf{16}$ of SO(10).  For the last nonabelian divergence, Eq. 
\myref{L52}, we explicitly write
out the contribution from the $\dot{Y}$ so that is takes the form
\bea
		 0 =  \text{Tr}(Q_X^L)C_{\mathcal{G}}^m + \text{Tr}\bigg\lbrack\eta Q_X^{PV} T_ aT_b\bigg\rbrack,
\eea
\noindent
where $C^m_{\mathcal{G}}$ is the Casimir of the representation of the matter fields.  If we consider fields from the set $S_{10}$, then this becomes
\bea -\text{Tr}(Q_X^L) &=&
\text{Tr}(Q_X^{PV}) = 2\text{Sum}\bigg\lbrack\eta
\bar{Q}^{PV}_X\bigg\rbrack\\  \eea 
\noindent 
The fields in $S_{10}$ contribute to Eq. \myref{L51} but not to Eq. \myref{L53} since we have restricted their $\gamma$ parameters to be $\gamma = \frac{1}{2}$. Their contribution to 
Eq. \myref{L51} is not an issue since we can simply include more fields in the other sets 
described in this section to cancel their contribution. Finally, the clone fields ensure that none of the sets described in this section contribute to other conditions.

\subsubsection{PV Fields for the Regulation of Abelian Divergences}\label{abel}
Here we satisfy the conditions Eqs. \myref{L61}--\myref{L68}. The
$\dot{Y}$ contribute here, and to cancel them we will need to
introduce fields with $\bar{q}_n\neq 0 $, which is different from all other
fields considered thus far. This would alter some of the expressions
we have used above, but we will not consider these alterations since
we will employ clone fields that cancel contributions to previously considered terms from the fields
introduced here. Specifically, we consider
\begin{itemize}
		\item $S_{11}$: A set of pairs of PV fields such that
                  the unprimed fields have the same abelian gauge
                  charges as the light fields (including U(1)$_X$),
                  $\alpha^P_{11} = \dot{\alpha}$, $\beta^P_{11} =
                  \dot{\beta}$, $q_n^{(11)} =
                  -q^{(L)}_n$, $\alpha^{\prime P} = \frac{1}{2}$,
                  $\beta^{\prime P} =Q_X^{\prime PV} = q^{\prime
                    (11)}_n= 0$, and positive signature and clone
                  fields with no U(1)$_a$ charges.
		\item$S_{12}$: A set of pairs of PV fields with no
                  $\beta$ parameters or modular weights and with
                  $\alpha^P_{12} = \alpha^{\prime P}_{12} = 1/2$,
                  $Q^{\prime X}_{(12)} = 0$, $Q^X_{(12)} = 4
                  Q^X_{(L)}$, and U(1)$_a$ charges $Q^a_{(12)} =
                  Q_{(L)}^a/\sqrt{2}$ and negative signature and clone
                  fields with no U(1)$_a$ charges.
\end{itemize}

These satisfy
{\footnotesize
\bea
		0 &=&
		 -S^{(L)}_{ab} + 2\text{Sum}\bigg\lbrack\eta_{11}Q^a_{(11)}Q^b_{(11)}\bigg\rbrack  + 
			2\text{Sum}\bigg\lbrack\eta_{12}Q^a_{(12)}Q^b_{(12)}\bigg\rbrack\\
		0 &=& -2\pi^2\delta_X + \text{Sum}\bigg\lbrack\eta_{11}Q^X_{(11)}Q^a_{(11)}Q^b_{(11)}\bigg\rbrack + \text{Sum}\bigg\lbrack\eta_{12}Q^X_{(12)}Q^a_{(12)}Q^b_{(12)}\bigg\rbrack\\
		0 &=& -\dot{\beta}S_a^{(L)} + R_a^{(L)} + \text{Sum}\bigg\lbrack\eta_{11}Q^a_{(11)}Q^X_{(11)}\beta_{11}\bigg\rbrack + \text{Sum}\bigg\lbrack\eta_{11}q^{(11)}_nQ^a_{(11)}Q^X_{(11)}\bigg\rbrack\\
		0 &=& - \left(\frac{1}{2} - \dot{\beta}\right)S^{(L)}_a + \text{Sum}\bigg\lbrack\eta_{11}Q^X_{(11)}Q^a_{(11)}\left(\alpha_{11}- \frac{1}{2}\right)\bigg\rbrack\\
		0 &=& -\frac{1}{2}S^{(L)}_a + R^{(L)}_a + \text{Sum}\bigg\lbrack\eta_{11}Q^X_{(11)}Q^a_{(11)}\left(\gamma_{11} - \frac{1}{2}\right)\bigg\rbrack + \text{Sum}\bigg\lbrack\eta_{11} Q^X_{(11)}Q^a_{(11)}q^{(11)}_n\bigg\rbrack\\
		0 &=& -\frac{1}{2}S_{ab}^{(L)} + R^{(L)}_{ab} + \text{Sum}\bigg\lbrack\eta_{11}Q^a_{(11)}Q^b_{(11)}\left(\left(\gamma_{11}-\frac{1}{2}\right) + q^{(11)}_n\right)\bigg\rbrack,
\eea}
\noindent
 where again a subscript or superscript (L) implies a trace over the
corresponding values of the light fields. Note that we have omitted some conditions that are automatically zero. There are also terms in the above that vanish for the choice of U(1) charges defined in this paper  but do not vanish for other choices. If
one substitutes the parameters of $S_{11}$ and $S_{12}$ as per the
discussion above, one sees that all the remaining conditions above are
satisfied.

\setcounter{equation}{0}
\subsection{The FIQS spectrum}\label{charges}

The FIQS model was described in
\cite{Ibanez:1987sn,Font:1988mm,fiqs,Casas:1988hb,Casas:1988se}. 
{The modular weights in this model are simple: the fields in the
$i$th untwisted sector have $q^i_n = \del^i_n$, and the twisted sector
fields have $q_n = {2\over3}$, except for the $Y^i$ with 
$q^i_n = \del^i_n + {2\over3}$.  Here  we}
will focus in particular on the U(1) charges of the low-energy matter
spectrum. The U(1) charge generators arising from the Cartan subalgebra
of the $E_8 \times E_8$ and the corresponding charges were worked out
in \cite{Casas:1988hb,Casas:1988se}.Table 2 of \cite{Font:1988mm}
lists the charges of the massless spectrum. However, the linear
combinations of generators given in
 \cite{fiqs} have a mixed anomaly:
\beq  		\Tr (Q_6Q_7Q_X) = 1296. \eeq
To avoid this, one should re-define $Q_6$ and $Q_7$. The fix is very simple:
\bea			Q_6^\prime &=& Q_6 - Q_7,\\
			Q_7^\prime &=& Q_6 + Q_7. \eea
Below we produce a table of the new charge designations.
{\scriptsize
\begin{center}
\begin{tabular}{c c c c c c c c c c c} 
 \hline
($n_1$, $n_3$) & Field & Rep & $Q_1$ & $Q_2$ & $Q_3$ & $Q_4$ & $Q_5$ & $Q_6^N$ & $Q_7^N$ & $X$\\ [0.5ex] 
 \hline
  untwisted & $Q_L$ &(3,2) & -6 & -6 & 0 & 0 & 0 & 0 & 0 & 0 \\
 	    & $u_L$ &  ($\bar{3}$,1)& 6 & 0 & 0 & -6 & 0 & 0 & 0 & 0 \\
 	    & $\bar{G}_1$ & (1,2) & 0 & 6 & 0 & 6 & 0 & 0 & 0 & 0 \\
 	    & $16^\prime$ & 1 & 0 & 0 & 0 & 0 & 0 & 0 & 0 & 9 \\

 \hline
 
 (0,0) &  $D_1$ & (3,1) & 0 & 4 & 0 & 0 & 0 & 4 & 4 & 4 \\
	& $\bar{G}_2$ &(1,2) & 6 & -2 & 0 & 0 & 0 & 4 & 4 & 4 \\
	&  $\bar{A}_1$& 1 & -3 & -2 & -3 & -3 & -3 & 4 & 4 & 4 \\
 	& $\bar{A}_2$& 1 & -3 & -2 & 3 & -3 & 3 & 4 & 4 & 4 \\
 	& $A_1$& 1 & -3 & -2 & -3 & 3 & 3 & 4 & 4 & 4 \\
	&  $A_2$ & 1 & -3 & -2 & 3 & 3 & -3 & 4 & 4 & 4 \\

 \hline
 
 (1,0) & $S_4$ &1 & 6 & 4 & 0 & 0 & -2 & 2 & 10 & 4 \\
	& $S_5$&1 & 6 & 4 & 0 & 0 & -2 & -4 & -8 & 4 \\
	& $S_6$&1 & 6 & 4 & 0 & 0 & -2 & 2 & -2 & -8 \\
	& $\bar{A}_3$&1 & -3 & -2 & -3 & -3 & 1 & 2 & 10 & 4 \\
	& $\bar{A}_4$&1 & -3 & -2 & -3 & -3 & 1 & -4 & -8 & 4 \\
	& $\bar{A}_5$&1 & -3 & -2 & -3 & -3 & 1 & 2 & -2 & -8 \\
	& $A_3$&1 & -3 & -2 & 3 & 3 & 1 & 2 & 10 & 4 \\
	& $A_4$ &1& -3 & -2 & 3 & 3 & 1 & -4 & -8 & 4 \\
	&  $A_5$&1 & -3 & -2 & 3 & 3 & 1 & 2 & -2 & -8 \\
 
\end{tabular}
 
\begin{tabular}{c c c c c c c c c c c} 
\hline
($n_1$, $n_3$) & Field & Rep & $Q_1$ & $Q_2$ & $Q_3$ & $Q_4$ & $Q_5$ & $Q_6^N$ & $Q_7^N$ & $X$\\ [0.5ex] 
  \hline
 
 (-1,0) & $S_7$&1 & 6 & 4 & 0 & 0 & 2 & 6 & -2 & 4 \\
 	&$S_8$ &1& 6 & 4 & 0 & 0 & 2 & 0 & 4 & -8 \\
 	&$S_9$ &1& 6 & 4 & 0 & 0 & 2 & -6 & -2 & 4 \\
 	&$\bar{A}_6$&1 & -3 & -2 & 3 & -3 & -1 & 6 & -2 & 4 \\
 	&$\bar{A}_7$&1 & -3 & -2 & 3 & -3 & -1 & 0 & 4 & -8 \\
 	&$\bar{A}_8$&1 & -3 & -2 & 3 & -3 & -1 & -6 & -2 & 4 \\
 	&$A_6$&1 & -3 & -2 & -3 & 3 & -1 & 6 & -2 & 4 \\
	& $A_7$ &1 & -3 & -2 & -3 & 3 & -1 & 0 & 4 & -8 \\
	& $A_8$&1 & -3 & -2 & -3 & 3 & -1 & -6 & -2 & 4 \\
 
  \hline
 
 (0,1) & $d_1$& $(\bar{3},1)$ & 0 & 0 & 0 & 2 & 2 & 0 & -8 & 4 \\
	& $F_1$ & (1,2) & 3 & 0 & -3 & -1 & -1 & 0 & -8 & 4 \\
 	&$\bar{A}_9$ &1 & 3 & 6 & 3 & -1 & -1 & 0 & -8 & 4 \\
	& $A_9$ &1 & 3 & -6 & 3 & -1 & -1 & 0 & -8 & 4 \\
	& $\bar{l}_1$&1 & -6 & 0 & 0 & -4 & 2 & 0 & -8 & 4 \\
	& $S_{10}$&1 & -6 & 0 & 0 & 2 & -4 & 0 & -8 & 4 \\
	
  \hline
 
 (1,1)	& $D_2$&(3,1) & 6 & 0 & 0 & 2 & 0 & -2 & -2 & 4 \\
	& $u_2$&($\bar{3}$,1) & 0 & 0 & 0 & -4 & 0 & -2 & -2 & 4 \\
	& $F_2$ &(1,2)& 3 & 0 & 3 & -1 & -3 & -2 & -2 & 4 \\
	& $F_3$&(1,2) & 3 & 0 & -3 & -1 & 3 & -2 & -2 & 4 \\
	& $S_1$&1 & -6 & 0 & 0 & 2 & 0 & 4 & 4 & -8 \\
	& $Y_1$&1 & -6 & 0 & 0 & 2 & 0 & -2 & -2 & 4 \\
	& $\bar{A}_{10}$&1 & 3 & 6 & -3 & -1 & -3 & -2 & -2 & 4 \\
	& $\bar{A}_{11}$&1 & 3 & 6 & 3 & -1 & 3 & -2 & -2 & 4 \\
	& $A_{10}$&1 & 3 & -6 & 3 & -1 & 3 & -2 & -2 & 4 \\
	& $A_{11}$&1 & 3 & -6 & -3 & -1 & -3 & -2 & -2 & 4 \\

  \hline
 
 (-1,1)	& $d_2$ & $(\bar{3},1)$ & 0 & 0 & 0 & 2 & -2 & -4 & 4 & 4 \\
 	&$F_4$ & (1,2) & 3 & 0 & 3 & -1 & 1 & -4 & 4 & 4 \\
 	&$\bar{A}_{12}$ & 1 & 3 & 6 & -3 & -1 & 1 & -4 & 4 & 4 \\
 	&$A_{12}$ &1 & 3 & -6 & -3 & -1 & 1 & -4 & 4 & 4 \\
 	&$\bar{l}_2$ &1 & -6 & 0 & 0 & -4 & -2 & -4 & 4 & 4 \\
 	&$S_{11}$&1 & -6 & 0 & 0 & 2 & 4 & -4 & 4 & 4 \\
 \hline
 
 (0,-1)	& $\bar{D}_1$ & $(\bar{3},1)$& -3 & 2 & -3 & 1 & 1 & 2 & -2 & 4 \\
 	&$D_3$&(3,1) & 3 & 2 & 3 & 1 & 1 & 2 & -2 & 4 \\
 	&$\bar{G}_3$&(1,2) & 0 & 2 & 0 & 4 & -2 & 2 & -2 & 4 \\
 	&$G_1$&(1,2) & 0 & 2 & 0 & -2 & 4 & 2 & -2 & 4 \\
 	&$S_2$&1 & 0 & -4 & 0 & -2 & -2 & -4 & 4 & -8 \\
 	&$Y_2$&1 & 0 & -4 & 0 & -2 & -2 & 2 & -2 & 4 \\
 	&$l_1$&1 & 0 & -4 & 0 & 4 & 4 & 2 & -2 & 4 \\
 	&$\bar{l}_3$&1 & 0 & 8 & 0 & -2 & -2 & 2 & -2 & 4 \\
 	&$\bar{A}_{13}$&1 & -9 & 2 & 3 & 1 & 1 & 2 & -2 & 4 \\
	& $A_{13}$&1 & 9 & 2 & -3 & 1 & 1 & 2 & -2 & 4 \\
 
\end{tabular}

\begin{tabular}{c c c c c c c c c c c} 
\hline
 ($n_1$, $n_3$) & Field & Rep & $Q_1$ & $Q_2$ & $Q_3$ & $Q_4$ & $Q_5$ & $Q_6^N$ & $Q_7^N$ & $X$\\ [0.5ex] 
   \hline
 
 (1,-1)	& $\bar{D}_2$ & $(\bar{3},1)$ & -3 & 2 & 3 & 1 & -1 & 0 & 4 & 4 \\
 	&$D_4$ & (3,1) & 3 & 2 & -3 & 1 & -1 & 0 & 4 & 4 \\
 	&$\bar{G}_4$&(1,2) & 0 & 2 & 0 & 4 & 2 & 0 & 4 & 4 \\
 	&$G_2$&(1,2) & 0 & 2 & 0 & -2 & -4 & 0 & 4 & 4 \\
 	&$S_3$&1 & 0 & -4 & 0 & -2 & 2 & 0 & -8 & -8 \\
 	&$Y_3$&1 & 0 & -4 & 0 & -2 & 2 & 0 & 4 & 4 \\
 	&$l_2 $&1 & 0 & -4 & 0 & 4 & -4 & 0 & 4 & 4 \\
 	&$\bar{l}_4$&1 & 0 & 8 & 0 & -2 & 2 & 0 & 4 & 4 \\
	& $\bar{A}_{14}$&1 & -9 & 2 & -3 & 1 & -1 & 0 & 4 & 4 \\
	& $A_{14}$&1 & 9 & 2 & 3 & 1 & -1 & 0 & 4 & 4 \\
 
  \hline

 (-1,-1)	& $G_3$& (1,2) & 0 & 2 & 0 & -2 & 0 & 4 & -8 & 4 \\
	& $G_4$& (1,2) & 0 & 2 & 0 & -2 & 0 & -2 & 10 & 4 \\
	& $G_5$& (1,2) & 0 & 2 & 0 & -2 & 0 & -2 & -2 & -8 \\
 	&$l_3 $&1 & 0 & -4 & 0 & 4 & 0 & -2 & 10 & 4 \\
 	&$l_4$&1 & 0 & -4 & 0 & 4 & 0 & 4 & -8 & 4 \\
 	&$l_5$&1 & 0 & -4 & 0 & 4 & 0 & -2 & -2 & -8 \\
\end{tabular}

\end{center}}

\subsection{Corrections to~\cite{gl}}\label{cors}

Equation (3.11) should read:

$$ \chiproj\Omega_W = W^{\alpha\beta\gamma}W_{\alpha\beta\gamma},
\qquad \chiproj\Omega_X = \Xc\Xa,\qquad\chiproj\Omega_{\rm Y M}
= \WaWa$$
\noindent

In Eqs. (3.6) and (5.2) the factor 1/24 in front of $\Omega_{\rm G B}$
should be removed.
\noindent

Equation (5.3) and the remainder of section 5 should read
$$ 8\pi^2b_{\rm spin} = 8\pi^2b + 1 = 31, \qquad\widetilde\Omega_f
= \Tr\eta\Del\ln\cM^2\Omega_f.\eqno(5.3)$$
\noindent
The results for the Gauss-Bonnet and Yang-Mills terms are 
well-established~\cite{prev} and result from the universality conditions
(2.3) and (B.7), as illustrated in the appendices. The
only other term in (5.2) that contains a chiral anomaly is
$\Omega_f$, which, using the set (4.11) of PV fields,
is a priori a product of the chiral superfields $\Xa,\;\ga$
and $\ga^n$.  We show in Appendix A that we may choose the PV parameters
such that 
$$ \chiproj\widetilde\Omega_f = 30\sum_n\gc_n\ga^n,\eqno(5.4)$$
\noindent
in agreement with the string calculation of [4].

The anomaly is canceled provided the Lagrangian for the dilaton $S,\S$
is specified by the coupling (2.5) and the K\"ahler potential
(2.9), or, equivalently, the linear superfield $L$ satisfies 
(1.3) and the GS term (1.3) is added to the Lagrangian.

\end{document}